\documentclass[fleqn]{article}
\usepackage{fourier}
\usepackage[psamsfonts]{eucal}         
\usepackage{graphicx}
\usepackage{cite}
\usepackage[cmex10]{amsmath}
\usepackage{amssymb}
\usepackage{bm}
\usepackage{psfrag}
\usepackage{subfigure}
\usepackage[usenames,dvipsnames]{color}
\usepackage[font=small]{caption}
\newcommand{\rmi}{\text{i}}
\newcommand{\x}{\boldsymbol{x}}
\newcommand{\q}{\boldsymbol{q}}

\newcommand{\kbf}{\boldsymbol{k}}
\newcommand{\bfk}{\boldsymbol{k}}

\newcommand{\ofx}{(\x)}
\newcommand{\ofk}{(\kbf)}
\newcommand{\etal}{\emph{et al. }}
\newcommand{\paperi}{\emph{Paper~I}}
\newcommand{\paperii}{\emph{Paper~II}}

\title{Effective Field Theory for Atom-Molecule Systems III: Dynamic Effects of a Feshbach Resonance on Bragg scattering from a Bose-Einstein Condensate}
\author{Catarina E. Sahlberg, R. J. Ballagh, C. W. Gardiner
\\[5mm] \emph{Jack Dodd Centre for Quantum Technology,} 
\\ \emph{Department of Physics, University of Otago,}
\\ \emph{ Dunedin, New Zealand}}

\begin{document}

\maketitle

\begin{abstract}
We present a theoretical model for Bragg scattering from a Bose-Einstein condensate (BEC) in the vicinity of a magnetic Feshbach resonance, using a two c-field formalism, one c-field for the atom and the other for a molecule formed of two atoms. We use this model to numerically simulate a recent experiment \cite{papp2008} investigating the effects of strong interactions on the Bragg spectrum from a $^{85}$Rb BEC. Results from these simulations are 
in very good quantitative agreement with the experimental results, confirming the importance of the resonance bound state in the dynamics of the condensate for fast experiments like Bragg scattering.
\end{abstract}

\section{Introduction}

Bose-Einstein condensates (BEC) with weak interparticle interactions have been, in many cases, successfully described using a \emph{pseudopotential} formulation, in which just two parameters are involved: the density and the s-wave scattering length. Furthermore, in a large proportion of situations, 
\emph{mean-field theory} can be used, leading to a description in terms of the Gross--Pitaevskii equation.  Arising from this success, a quest for a tunable and possibly large interaction strength began,  leading to the study of systems in which the scattering length was the result of a Feshbach resonance, whose use made it possible to tune the interatomic interaction of a BEC over a wide range.  

Using Feshbach resonances, it has become possible to investigate condensates with strong interparticle interactions.  This has been done both experimentally and theoretically with quite wide success, even though there are reasons to question the validity of the standard theoretical procedures at some of the interaction strengths used.  The pseudopotential and mean-field theory methods are the result of a perturbation treatment, which must definitely fail for sufficiently high interaction strengths and densities.

With this in mind, in \paperi\ \cite{sahlberg1} and \paperii\ \cite{sahlberg2} we introduced a more careful treatment of  interactions mediated by weakly bound molecular states, such as arise in a Feshbach resonance.  This was done by introducing a molecular field, whose interaction constants can be determined phenomenologically from scattering length and binding energy data.
In our treatment, the interaction constants are relatively weak, but nevertheless reproduce many of the results of a simple pseudopotential method, especially for static properties, such as the condensate shape. Building on this, in \paperii\ we formulated a Bogoliubov description, and showed that there are changes in the excitation spectrum at higher energies.  Using this description, we found modifications in the Bragg scattering spectrum from a homogeneous infinite condensate very similar to those which were experimentally found by
 Papp \emph{et al.} \cite{papp2008}

What this means is that, by treating the dynamics behind the change in interaction strength in the Feshbach resonance, we find that a mean-field treatment of a strongly interacting system is still very much applicable, provided that there are two mean fields, one for the atoms and one for the molecules.

In this paper we continue the work presented in our two previous papers, and investigate a realistic system. We apply the formalism of \paperi\ and \paperii\ to the specific case of a recent experiment by Papp \etal \cite{papp2008}, in which the excitation spectrum of a Bose-Einstein condensate of $^{85}$Rb was measured using Bragg scattering, near the Feshbach resonance at 155 G. This experiment was deliberately designed to explore a region of parameter space in which the perturbation theory would not be expected to be valid.  And indeed, by tuning the scattering length to large values, they found significant deviations from the Bragg scattering behaviour predicted by the simplest perturbative and mean-field theories.

The results of our work are very satisfactory.  Although the experiment was not designed to test this kind of theory, and thus some significant parameters are hard to estimate,  we obtain quantitative agreement with their experimental results, with no fitted parameters. 
\subsection{Properties of the Bragg spectrum}
\label{sec:IntroProperties}
The excitation spectrum of a homogeneous condensate for large momentum transfer is given by the sum of the kinetic energy and the chemical potential of the condensate,
\begin{equation}
\label{eq:simpleshift}
\hbar\omega(k) = \frac{\hbar^2k^2}{2m} + \frac{4\pi\hbar^2na_s}{m},
\end{equation} 
where $k$ is the photon momentum, $n$ is the density of the condensate and $a_s$ is the s-wave scattering length. In the case of an inhomogeneous condensate, for example a condensate in a
harmonic trapping potential, it was found by Stenger \etal  \cite{stenger1999} that this formula
can be used provided $n$ is interpreted as the \emph{density-weighted density} of the condensate. In their work, the use of the density-weighted density was theoretically justified by using a local density description of the condensate. Blakie \etal \cite{blakie2000,blakie2002} simulated the system using the Gross-Pitaevskii equation, and confirmed the basic validity of this approximation. In the region where (\ref{eq:simpleshift}) is valid, it is equivalent to the Bogoliubov excitation spectrum in the limit of large $k$. 

The prediction (\ref{eq:simpleshift}) is expected to be valid as long as the condensate is dilute ($na_s^3\ll 1$), the excitation is in the free-particle regime ($k\xi\gg 1$, where $\xi = ({8\pi na_s})^{-1/2}$ is the condensate healing length) and the scattering amplitude is momentum independent ($ka_s\ll 1$). The aim of the experiment of Ref. \cite{papp2008} was to investigate the properties of the Bragg spectrum in a region where the scattering length is large. This means that the condensate interactions cannot be treated as mean-field ($\sqrt{8\pi n a_s^3} \sim 0.5$), the excitations are not clearly particle-like ($k\xi\sim2$), and the scattering amplitude is not clearly momentum independent ($ka_s \sim0.8$).

\subsection{Experimental results and issues}
In the experiment of Ref. \cite{papp2008}, the shifts of the Bragg spectra for large scattering lengths showed a significant deviation from the theoretical predictions based on (\ref{eq:simpleshift}). The experimental results were also compared to theoretical predictions outlined in detail in \cite{ronen2009}, which are not in agreement with the experimental data.

Another theoretical model is presented by Kinnunen \etal \cite{kinnunen2009}, who studied Bragg spectroscopy from a uniform, strongly interacting $^{85}$Rb condensate using time-dependent Hartree-Fock-Bogoliubov theory. They took into account the momentum dependent scattering amplitude, but found only qualitative agreement with the experiment.  

\subsection{Interpreting the Experiment}%
\begin{figure}[t]\begin{center}
\begin{psfrags}%
\psfragscanon%
\psfrag{s01}[t][t]{\color[rgb]{0,0,0}\setlength{\tabcolsep}{0pt}\begin{tabular}{c}Scattering length $a_s$ [$a_0$]\end{tabular}}%
\psfrag{s02}[b][b]{\color[rgb]{0,0,0}\setlength{\tabcolsep}{0pt}\begin{tabular}{c}Frequency [kHz]\end{tabular}}%
\psfrag{s05}[l][l]{\color[rgb]{0,0,0}\setlength{\tabcolsep}{0pt}\begin{tabular}{l}Binding frequency ${f}_b$\end{tabular}}%
\psfrag{s06}[l][l]{\color[rgb]{0,0,0}\setlength{\tabcolsep}{0pt}\begin{tabular}{l}Range of Bragg frequencies ${f}_{\mbox{\footnotesize{Bragg}}}$\end{tabular}}%
\psfrag{s07}[l][l]{\color[rgb]{0,0,0}\setlength{\tabcolsep}{0pt}\begin{tabular}{l}Inverse pulse length ${f}_{\mbox{\footnotesize{pulse}}}$\end{tabular}}%
\psfrag{x01}[t][t]{0}%
\psfrag{x02}[t][t]{0.2}%
\psfrag{x03}[t][t]{0.4}%
\psfrag{x04}[t][t]{0.6}%
\psfrag{x05}[t][t]{0.8}%
\psfrag{x06}[t][t]{1}%
\psfrag{x07}[t][t]{200}%
\psfrag{x08}[t][t]{400}%
\psfrag{x09}[t][t]{600}%
\psfrag{x10}[t][t]{800}%
\psfrag{v01}[r][r]{0}%
\psfrag{v02}[r][r]{0.1}%
\psfrag{v03}[r][r]{0.2}%
\psfrag{v04}[r][r]{0.3}%
\psfrag{v05}[r][r]{0.4}%
\psfrag{v06}[r][r]{0.5}%
\psfrag{v07}[r][r]{0.6}%
\psfrag{v08}[r][r]{0.7}%
\psfrag{v09}[r][r]{0.8}%
\psfrag{v10}[r][r]{0.9}%
\psfrag{v11}[r][r]{1}%
\psfrag{v12}[r][r]{$10^{0}$}%
\psfrag{v13}[r][r]{$10^{1}$}%
\psfrag{v14}[r][r]{$10^{2}$}%
\psfrag{v15}[r][r]{$10^{3}$}%
\psfrag{v16}[r][r]{$10^{4}$}%
\resizebox{8.5cm}{!}{\includegraphics{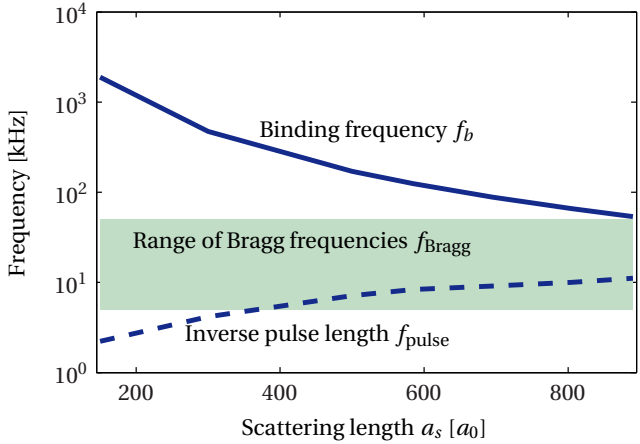}}%
\end{psfrags}%
\caption{Frequency scales in the experiment \cite{papp2008}.}
\label{fig:freq}
\end{center}\end{figure}

\label{Interpreting the Experiment}%
There are several issues that complicate the analysis of the experiment of Ref. \cite{papp2008}. We will address the most important of these in the following:
\begin{enumerate}
\item \emph{Initial state} : 
The nonlinear effects that were investigated in the experiment are more pronounced the less dilute the condensate is. The density of the initial state in the experiment is therefore enhanced by a series of ramps of the scattering length. Creating an initial state in this way makes the experimental procedure even more complex and takes the system further away from the ideal case studied theoretically by \cite{ronen2009} and \cite{kinnunen2009}.

Furthermore, the initial state parameters are not explicitly defined, which makes analysis of the line shift result difficult, since this depends on the properties of the condensate at the onset of the Bragg pulse. 

\item\emph {Inhomogeneity} : 
The trapped condensate is spatially inhomogeneous, and also strongly time varying, because of both three-body losses and condensate expansion in the breathing modes, and even as a result of the Bragg scattering process itself (as we shall see in section \ref{sec:ResultsBehaviour}).  Papp \etal measure and use \emph{space averaged} densities, rather than the \emph{density weighted} densities, which (as we noted above) are more appropriate  when comparing with results for a homogeneous condensate.  In addition to this, they also average densities over the duration of the experiment.

The problem with using the space-averaged density is that unless the condensate has a clearly defined volume, the space-averaged density cannot be accurately determined. In an experiment such as that of \cite{papp2008}, the volume is not easily determined and has to be approximated in one way or another. 

In \cite{papp2008}, the time- and space-averaged density was determined by assuming that the density profile of the condensate is given by a Thomas-Fermi profile with a width given by a variational solution to the Gross-Pitaevskii equation (GPE). The variational model, outlined in \cite{perez-garcia1997} is, however, not an accurate representation of an exact solution of the GPE, which even in three dimensions, is not very difficult to find numerically. 

Furthermore, it is not clear how accurate a description of the shape of the condensate is given in this case by a Thomas-Fermi profile. As we shall see in section \ref{sec:ResultsBehaviour}, in our simulations the shape of the condensate is very different from that given by the Thomas-Fermi approximation, both before and during the application of the Bragg pulse.
 
\item\emph{Variable pulse length and intensity} : 
The condensate density varies more rapidly in time as a result of Bragg scattering at larger scattering lengths. By introducing the condition that the density of the condensate cannot change by more than 30\% during the Bragg pulse, based on predictions from the variational model, the experiment is forced to use progressively shorter Bragg pulses for larger scattering length. To make sure that roughly the same quantity is scattered out each time, the intensity of the pulses is appropriately increased. 

In our calculations we find that the processes involved are sensitive to the duration and intensity of the Bragg pulse, because the condensate expands, because there are three-body losses, and because the spectroscopic resolution improves for longer pulses.  It is therefore important to reproduce the experimental parameters as faithfully as possible.  Unfortunately, neither duration nor the intensity of the Bragg pulse are explicitly stated in \cite{papp2008}, so we have inferred their values from the spectra and the limits on the number of Bragg-scattered atoms.

\item\emph{Time scales} :
Bragg scattering is a fast process and it is therefore important to consider the other time scales associated with the experiment; if other processes occur on a time scale similar to that of the Bragg scattering, it is likely that those processes are important for the dynamics of the condensate.

There are three time scales that are relevant in this type of experiment, shown in Fig.\,\ref{fig:freq} in terms of their corresponding frequencies: The frequency of the applied Bragg pulse, the inverse of the pulse duration and the binding frequency of the bound state in the Feshbach resonance. For large scattering lengths and short Bragg pulses, these frequencies are comparable, and it is therefore very probable that the bound state dynamics become important to the overall dynamics of the experiment.
\end{enumerate}

\section{Formalism}
The formalism of \paperi\ and \paperii\ proceeds in brief as follows:
To model the Feshbach resonance bound state, we add an additional field, corresponding to a bound atom pair (refered to as a ``molecule''), to the usual Hamiltonian for a trapped system of interacting Bosons.
The equations of motion for the atom field $\psi$ and molecule field $\phi$ in the resulting c-field model are given by
\begin{eqnarray}
\label{eq:dPsi_dt}
\rmi\hbar\frac{\partial\psi\ofx}{\partial t} &=&  -\frac{\hbar^2\nabla^2}{2m}\psi\ofx  + \mathcal{P}_a\left\{V_a\ofx\psi\ofx + U_{aa}|\psi\ofx|^2\psi\ofx + g\psi^*\ofx\phi\ofx\right\} \nonumber \\ 
&& -\textrm{i}\gamma\left(|\psi\ofx|^2+2|\phi\ofx|^2\right)^2\psi\ofx,\\
\label{eq:dPhi_dt} 
\rmi\hbar\frac{\partial\phi\ofx}{\partial t} &=& -\frac{\hbar^2\nabla^2}{4m}\phi\ofx + \mathcal{P}_m\left\{ \left(\varepsilon+  V_m\ofx\right)\phi\ofx + \frac{g}{2}\psi^2\ofx\right\},
\end{eqnarray}
where $U_{aa}=4\pi\hbar^2a_{bg}/m$ is the background interaction strength, $V_a$ and $V_m$ are the external trapping potential for the atoms and molecules respectively. The last term in (\ref{eq:dPsi_dt}) is added to account for losses from the condensate due to three-body recombination events \cite{norrie2006}; we discuss this more extensively in Sect.\,\ref{sec:SimParameters3body}.

\subsection{Projectors}
$\mathcal{P}_a$ and $\mathcal{P}_m$ are the atom and molecule projectors that restrict the wavefunctions 
to the low energy subspace below the momentum cutoff,
\begin{eqnarray}
\label{eq:Pa}
\mathcal{P}_a &=& \Theta \left(\left(\frac{k_x^2}{k_{x,\text{cut}}^2} + \frac{k_y^2}{k_{y,\text{cut}}^2} + \frac{k_z^2}{k_{z,\text{cut}}^2} \right)- 1\right)  
\\ \label{eq:Pm}
\mathcal{P}_m &=&  \Theta \left(\frac{1}{4}\left(\frac{k_x^2}{k_{x,\text{cut}}^2} + \frac{k_y^2}{k_{y,\text{cut}}^2} + \frac{k_z^2}{k_{z,\text{cut}}^2} \right)- 1\right)
\end{eqnarray}
where $\Theta$ is the Heaviside step function, and the momentum space cutoff in the $j$-th direction is given by
\begin{equation}
k_{j,\text{cut}} = \frac{\max{(k_j)} }{2} .
\end{equation}
These projectors arise from the simulation grid, and they are defined in order to avoid aliasing in our simulations \cite{blakie2008}. However, as discussed in \paperi, it is necessary to include the momentum space cutoff for two other reasons: in order for a pseudopotential treatment of the interaction to be valid and in order to avoid excessive quantities of the noise being added in the c-field method.

\subsection{Parameters}
The parameters $g$ and $\varepsilon$ are the coupling strength and detuning in the Feshbach resonance respectively. As shown in \paperi, they are in our formalism given by 
\begin{eqnarray}
\label{eq:epsilon}
\varepsilon &=&{\hbar^2\alpha^2\over 2m}
\frac{\left(\pi-2\Lambda a_s\right)
\left (1-\frac{2\Lambda a_{bg}}{\pi} t(\frac{\alpha}{\Lambda}) \right)}
{\Lambda a_s(1+t(\frac{\alpha}{\Lambda}))-\pi} 
\,,\\
\label{eq:g2}
g^2
&=& {8\pi\hbar^4\alpha^2\over m^2}
{\left(a_{bg}(\pi-2\Lambda a_s) -\pi a_s\right)
\left(1-\frac{2\Lambda a_{bg}}{\pi} t(\frac{\alpha}{\Lambda})\right)
\over
2\Lambda a_s \left(1+t(\frac{\alpha}{\Lambda})\right) -\pi}\, .
\end{eqnarray}
where $t(x)=x-\arctan{1/x}$, and $\hbar^2\alpha^2/m$ is the molecular binding energy 
corresponding to the s-wave scattering length $a_s$ \cite{sahlberg1}. The parameter $\Lambda$ is the renormalization factor, determined by the momentum space cutoffs. 

The effect of the Bragg field on the condensate is included by making the following substitutions in the equations of motion:
\begin{eqnarray}
V_a &\rightarrow& V_a +V_{\text{opt}}, \qquad
V_m \rightarrow V_m + 2V_{\text{opt}}
\end{eqnarray}
where 
\begin{equation}
\label{eq:braggpulse}
V_{\text{opt}} = V_0\cos{(\x\cdot\q -\omega t)},
\end{equation}
where $\q$ and $\omega$ are the wavevector and the frequency of the Bragg pulse respectively \cite{blakie2000,blakie2002}, and $V_0$ is the amplitude of the optical potential, given in terms of the Rabi frequency $\Omega$ and the excited state detuning $\Delta$,
\begin{equation}
V_0 = \frac{\hbar\Omega^2}{2\Delta}.
\end{equation}
The optical potential for the molecule is chosen to be twice that of the atom on the assumption that the atoms in the molecule are very weakly bound, and for these purposes behave almost independently.

\subsection{Renormalization factor}
We tend to refer to the parameter $\Lambda$ as the momentum space cutoff. However, in reality this is only true in the special case of isotropic cutoffs. 
The relationship between the renormalization constant $\Lambda$ and the momentum space cutoffs $k_{x,\text{cut}}$, $k_{y,\text{cut}}$ and $k_{z,\text{cut}}$, is given by
\begin{equation}
\label{eq:renormalization}
4\pi\Lambda = \int_V{\frac{d\kbf}{k^2}}, 
\end{equation}
where $V$ is the ellipsoidal volume spanned by the momentum space vectors, corresponding to the projectors (\ref{eq:Pa}, \ref{eq:Pm}).

In the simplest case, the momentum space cutoff is the same in all directions, and the volume of the populated low energy subspace is spherical so that evaluating (\ref{eq:renormalization}) gives
\begin{equation}
\Lambda  =  k_{R,\text{cut}},
\end{equation}
where $k_{R,\text{cut}}$ is the value of the isotropic cutoff. In the case of an anisotropic cutoff, as in the case for the numerical calculations in this paper, the exact value of $\Lambda$ needs to be evaluated using equation (\ref{eq:renormalization}). More details are given in Appendix \ref{sec:AppRenormalization}.

\section{Simulations}
We simulate the experiment of Ref. \cite{papp2008} by numerically solving the equations of motion (\ref{eq:dPsi_dt}) and (\ref{eq:dPhi_dt}) in three dimensions. To model the effects of quantum fluctuations in c-field theory, the wavefunctions in (\ref{eq:dPsi_dt}) and (\ref{eq:dPhi_dt}) have a random amplitude added to the initial states, corresponding to half a virtual particle per mode \cite{blakie2008}. 


\subsection{Momentum space truncation}
The trapping potential in the experiment of \cite{papp2008} is cigar-shaped, with an aspect ratio of $1:46$. This, along with the fact that the Bragg pulse is applied in the axial direction, and the Bragg momentum is relatively large, leads to a system that is computationally demanding. To include all the relevant physics, and at the same time ensuring that the c-field methods are still valid, and that the system is still computationally tractable, we make a truncation of the momentum space, neglecting all the modes that do not make a significant contribution to the dynamics of the system. 
This procedure, which involves dividing momentum space into bands, each centered around one of the Bragg orders, expresses the wavefunctions $\psi$ and $\phi$ as
\begin{align}
\psi\ofx &= \sum_n \psi_n\ofx e^{inQx} ,\\
\phi\ofx &= \sum_n \phi_n\ofx e^{inQx} ,
\end{align}
where $\psi_n$ and $\phi_n$ are the Fourier transforms of the momentum space wavefunction in the band centred around $nQ$. The details of the procedure are given in Appendix \ref{sec:AppTruncation}.

We find that only the four momentum bands
corresponding  to the orders  $n = -1,0,1,2$ acquire sufficient population to affect the simulation. For each band there will be an atom wavefunction $\psi_n$ and a molecule wavefunction $\phi_n$. The equations of motion for the atom wavefunctions are given by
\begin{eqnarray}
\label{eq:psi_-1}
\rmi\hbar\frac{\partial\psi_{-1}}{\partial t} &=&-\frac{\hbar^2\tilde\nabla_{-1}^{2}}{2m}{\psi_{-1}} + \mathcal{P}_a\left\{V_{a}\psi_{-1} +\frac{V_0}{2}\psi_{0} e^{i\omega t} \right. \nonumber \\
&&+ U_{aa}\left(A_0\psi_{-1}  + A_{-1}\psi_{0} + A_{-2}\psi_{1}+ A_{-3}\psi_{2}  \right) +g\left(\psi_{0}^*\phi_{-1} +\psi_1^*\phi_0  + \psi_{2}^*\phi_{1}  \right)    \nonumber\\
&& \left.  - i\gamma\left(C_0\psi_{-1}  + C_{-1}\psi_{0} + C_{-2}\psi_{1} + C_{-3}\psi_{2} \right) \right\} \\
\label{eq:psi_0}
\rmi\hbar\frac{\partial\psi_0}{\partial t} &=& -\frac{\hbar^2\tilde\nabla^{2}_0}{2m}{\psi_0} + \mathcal{P}_a\left\{V_{a}\psi_0 +\frac{V_0}{2}\left(\psi_{-1} e^{-i\omega t} + \psi_1 e^{i\omega t} \right) \right. \nonumber \\
&& + U_{aa}\left(A_{1}\psi_{-1} + A_0\psi_{0} + A_{-1}\psi_{1} + A_{-2}\psi_{2} \right) +g\left(\psi_{-1}^*\phi_{-1} +\psi_0^*\phi_0 + \psi_1^* \phi_1   + \psi_2^* \phi_2  \right)\nonumber\\
&& \left.   - i\gamma\left(C_{1}\psi_{-1}  + C_0\psi_{0} +  C_{-1}\psi_{1} +C_{-2}\psi_{2}\right) \right\} \\
\label{eq:psi_1}
\rmi\hbar\frac{\partial\psi_1}{\partial t} &=&-\frac{\hbar^2\tilde\nabla_1^{2}}{2m}{\psi_1} + \mathcal{P}_a\left\{V_{a}\psi_1 +\frac{V_0}{2}\left(\psi_{0} e^{-i\omega t} + \psi_2 e^{i\omega t} \right)\right. \nonumber \\
&&+ U_{aa}\left(A_{2}\psi_{-1}+ A_{1}\psi_{0} + A_0\psi_{1}  + A_{-1}\psi_{2} \right) \nonumber\\
&&  \left. +g\left(\psi_{1}^*\phi_{2} +\psi_0^*\phi_1 + \psi_{-1}^* \phi_0  \right) - i\gamma\left(C_0\psi_{1}  + C_{1}\psi_{0} + C_{2}\psi_{-1} \right) \right\} \\
\label{eq:psi_2}
\rmi\hbar\frac{\partial\psi_2}{\partial t} &=& -\frac{\hbar^2\tilde\nabla_2^{2}}{2m}{\psi_2} + \mathcal{P}_a\left\{V_{a}\psi_2 +\frac{V_0}{2}\psi_{1} e^{-i\omega t} + U_{aa}\left(A_3\psi_{-1} + A_{2}\psi_{0} + A_{1}\psi_{1} + A_{0}\psi_{2} \right) \right. \nonumber\\
&& \left. +g\left(\psi_{-1}^*\phi_{1} +\psi_0^*\phi_2 \right) - i\gamma\left(C_3\psi_{-1}  + C_{2}\psi_{0} + C_{1}\psi_{1} + C_{0}\psi_{2} \right) \right\}, 
\end{eqnarray}
where for brevity we have suppressed the spatial and temporal dependence, and where
\begin{equation}
\tilde\nabla_n^2 = \nabla^2 + \rmi2nQ\frac{\partial}{\partial x} - n^2Q^2 ,
\end{equation}
and the factors $A_n$, $B_n$ and $C_n$ are given in Appendix \ref{sec:AppTruncation}.

Similarly, the equations of motion for the molecule wavefunctions become
\begin{eqnarray}
\label{eq:phi_-1}
\rmi\hbar\frac{\partial\phi_{-1}}{\partial t} &=& -\frac{\hbar^2\tilde\nabla_{-1}^{2}}{4m}{\phi_{-1}} + \mathcal{P}_m\left\{V_{m}\phi_{-1} +V_0\phi_{0} e^{i\omega t} + {g}\psi_{-1}\psi_{0}  \right\} \\
\label{eq:phi_0}
\rmi\hbar\frac{\partial\phi_{0}}{\partial t} &=& -\frac{\hbar^2\tilde\nabla_0^{2}}{4m}{\phi_{0}} + \mathcal{P}_m\left\{V_{m}\phi_{0} +{V_0}\left(\phi_{-1} e^{-i\omega t} + \phi_1 e^{i\omega t} \right) \right.\nonumber\\
&&+ \left.\frac{g}{2}\left(2\psi_{-1}\psi_{1} + \psi_0^2 \right)\right\} \\
\label{eq:phi_1}
\rmi\hbar\frac{\partial\phi_{1}}{\partial t} &=& -\frac{\hbar^2\tilde\nabla_1^{2}}{4m}{\phi_{1}} + \mathcal{P}_m\left\{V_{m}\phi_{1} +{V_0}\left(\phi_{0}  e^{-i\omega t} + \phi_2  e^{i\omega t} \right) \right.\nonumber\\
&&\left.+ g\left(\psi_{-1}\psi_{2} + \psi_0\psi_1 \right)\right\} \\
\label{eq:phi_2}
\rmi\hbar\frac{\partial\phi_{2}}{\partial t} &=& -\frac{\hbar^2\tilde\nabla_2^{2}}{4m}{\phi_{2}} + \mathcal{P}_m\left\{V_{m}\phi_{2} +{V_0}\phi_{1} e^{-i\omega t}  + \frac{g}{2}\left(2\psi_0\psi_2+ \psi_1^2 \right)\right\}. 
\end{eqnarray}

The projectors $\mathcal{P}_a$ and $\mathcal{P}_m$ are defined as in (\ref{eq:Pa}, \ref{eq:Pm}) with the $k_x$-directional cutoff now given by
\begin{equation}
k_{x,\text{cut}} = \frac{\Delta k}{4},
\end{equation}
where $\Delta k$ is the width of each momentum space band. The projectors are the same for all the wavefunctions $\psi_n$ and $\phi_n$; for each band the projectors are ellipsoids centered around the midpoint at $k_y=k_z=0$ and $k_x=nQ$. The full wavefunctions, $\psi$ and $\phi$, are thus projected onto four disjoint regions in momentum space (see Fig.\,\ref{fig:Projector} in Appendix\,\ref{sec:AppTruncation}). The band width $\Delta k$ is chosen as a compromise between two factors: it needs to be large enough to include as much as possible of the momentum space wavefunction, but at the same time small enough to not include too much of the initial noise. It is also important that the individual bands are not overlapping.

\begin{table}[t]
\caption{\label{tab:density}Experimental average densities (column 3) for different scattering lengths (column 1). The data for the shift (column 2) are taken from Fig. 3(a) in \cite{papp2008}. Column 4 lists the approximate ranges of the average densities, where the values in brackets are our estimates. Column 5 shows the time averages of the density-weighted density that we use in our numerical calculations.}\begin{center}\small
\begin{tabular}{c|c|c|c|c}\hline
$a_s$ &$f_{\text{shift}}$&Time- and space-average&Range& Time-average density-weighted \\
$[a_0]$&[kHz]& density $[10^{19}$ m$^{-3}]$& $[10^{19}$ m$^{-3}]$ & density $[10^{19}$ m$^{-3}]$ \\
\hline
150&0.9&7.6&7.6&10.857 \\
300&1.7&7.2&7.2 &10.286\\
500&2.9&7.3&7.1--7.6& 10.429\\
585&3.0&6.5&(5.6 -- 7.4)&\hspace{0.5em}9.286\\
695&3.8&6.9&(6.2 -- 7.6)&\hspace{0.5em}9.857\\
805&3.9&6.1&4.9--7.4& \hspace{0.5em}8.714\\
890&4.6&6.5&(4.8 -- 8.2)&\hspace{0.5em}9.286\\
\hline
\end{tabular}
\end{center}
\end{table}

\subsection{Simulation parameters}
In choosing the parameters for our simulation we follow the experimental setup as closely as possible, and model a condensate of 40,000 $^{85}$Rb atoms in a trap with cylindrical symmetry and an aspect ratio of $46.2$ ($\nu_z = 2.9$Hz, $\nu_r = 134$Hz). The scattering length $a_s$ ranges from $150a_0$ to $890a_0$, and we use relationship between the scattering length and the parameters $g$ and $\varepsilon$ derived in \paperi\ and given by equations (\ref{eq:epsilon}) and (\ref{eq:g2}).

\subsubsection{Initial state}
\label{sec:SimInitial}
In the experiment an initial condensate was created with a scattering length of 150$a_0$. The scattering length was then ramped to an unspecified low value, exciting the large amplitude breathing modes in the condensate. At the inner radial turning point of the breathing mode oscillation, the scattering length was ramped up to the desired value, and the Bragg pulse was applied. Through this process the condensate becomes much denser, making the nonlinear effects on the Bragg spectra more clearly visible. However, it is hard to know exactly what the initial state at the commencement of the Bragg pulse is; had the condensate not been compressed in this way, the initial state would have been clearly defined.

In our simulations we create the initial state for the Bragg spectroscopy by performing the following steps:
\begin{enumerate}
\item We set the scattering length to a small, arbitrary value  $a_\text{init}$, typically of the order of a few $a_0$.
\item We numerically solve the time-independent equivalents of the equations of motion (\ref{eq:dPsi_dt}) and (\ref{eq:dPhi_dt}) for this scattering length, using the Thomas-Fermi equations that we derived in \paperii\ as the starting position. 
\item We quickly ramp the scattering length up to the value of interest, in the range between 150$a_0$ and 890$a_0$. The speed of the ramp never exceeds $\dot{a_s}/a_s < 0.25\hbar/ma_s^2$.

\item We then apply the Bragg pulse while continuing to run the simulation,  and calculate the resulting time-average of the density weighted density for the duration of the Bragg pulse. In our simulations, the density weighted density $\tilde{n}(t)$ is given by 
\begin{eqnarray}
\label{eq:spaceaverage}
\tilde{n}(t)&=&\frac{1}{N(t)}\int{d\x[|\psi(\x,t)|^2+2|\phi(\x,t)|^2]^2},
\\ \label{atomTotalNumber}
N(t)&=&\int{d\x[|\psi(\x,t)|^2+2|\phi(\x,t)|^2]}.
\end{eqnarray}
Here $N(t)$ and $\tilde{n}(t)$ involve  ``effective numbers of atoms'', counting each molecule as two atoms, corresonding to what would in practice be measured in an experiment.

\item We adjust $a_\text{init}$ appropriately, and redo steps 1-4 until the time-averaged density weighted density obtained matches that of the experiment.
\end{enumerate}

The experimental space- and time-averaged density is inferred from the predictions of the line shift (Fig. 3(a) in \cite{papp2008}). This varies from $7.6\times10^{13}$cm$^{-3}$ for $150a_0$ to between approximately $4.8$ and $8.2\times10^{13}$cm$^{-3}$ for $890a_0$, see Tab.\,\ref{tab:density}. We relate the experimental \emph{space-averaged} density to our \emph{density-weighted} density by noting that the space-averaged density for a Thomas--Fermi profile is $0.7$ times the density averaged density. The same factor is not necessarily right for other profile shapes, and as we shall see, the condensate in our simulations is quite far from Thomas--Fermi shaped.  However, we believe that this nonetheless gives us the best estimate of the density used in \cite{papp2008} that we can reasonably expect to get, since it corresponds to the procedure used in the experiment to estimate the space-averaged density.

\subsubsection{Bragg pulse}
The Bragg pulse, modelled by (\ref{eq:braggpulse}) and assumed to be square, is applied at the start of the simulation with a 
wavenumber of $k = 4\pi/780$ nm in the axial direction of the condensate. The pulse durations are not explicitly stated in \cite{papp2008}, but can be inferred from the data for the widths of the spectra (Fig.\,3(b) in \cite{papp2008}), where the contribution from the pulse duration will be inversely proportional to the pulse length as $\Delta\omega = 0.36/t_{\rm{pulse}}$ (where the number 0.36 comes from the rms width of a Gaussian fit to the Fourier transform of a square Bragg pulse), see Tab.\,\ref{tab:App} in Appendix\,\ref{sec:parameters}. 

{\begin{table}[bt]
\caption{\label{tab:pulse} Properties of the Bragg pulse used in
our simulations for different scattering lengths. The values of the duration are inferred from the spectral widths in \cite{papp2008}.}
\begin{center}\small
\begin{tabular}{c|c|c|c}\hline
Scattering length $a_s$ &Duration $t$ &Amplitude $V_0$&Scattered fraction \\
$[a_0]$ & [ms]& [$h$ kHz] &  \\
\hline
150&0.45& 0.13 & 6.2\% \\
300&0.24& 0.40 & 6.7\%\\
500&0.14& 0.67 & 6.5\%\\
585&0.12& 0.80 & 7.5\%\\
695&0.11& 0.80 & 6.4\%\\
805&0.10& 0.94& 7.1\%\\
890&0.09& 1.07& 8.0\%\\
\hline
\end{tabular}
\end{center}
\end{table}}

Thus, the pulse duration, and therefore also the simulation time, ranges from $0.09$ ms for $890a_0$ to $0.45$ ms for $150a_0$, see Tab.\,\ref{tab:pulse}. In Tab.\,\ref{tab:pulse}, we have also listed the Bragg pulse amplitudes $V_0$ for the different values of the scattering length. The intensities of the Bragg pulse are not stated in \cite{papp2008}, but as in the experiment we have chosen it so that we always have between 5\% and 10\% of the condensate being scattered, see Tab.\,\ref{tab:pulse}.

\subsubsection{Three-body loss}%
\label{sec:SimParameters3body}
The rate of particle loss from the condensate arising from three-body recombination events varies with the scattering length, approximately proportionally to $a_s^4$ \cite{braaten2006}, and is also extremely sensitive to the density. There are no exact values of the three-body loss coefficient ($\gamma$ in (\ref{eq:dPsi_dt})) available; here we have used the theoretical values given by Braaten \etal \cite{braaten2007}, which qualitatively agreed with previous experimental data from Roberts \etal \cite{roberts2000}. However, as is noted in \cite{braaten2007} these values are highly uncertain.

The way in which to include three-body loss in a c-field formalism was originally developed by Norrie \etal \cite{norrie2006}. This treatment includes a stochastic term in the equations of motion, but Norrie shows that this term can in most cases be neglected to a good order of approximation. In (\ref{eq:dPsi_dt}), we have used this approximate form, and have  extended it phenomenologically to include the molecule population as well. 
This extension is quite simple minded.  We include all of the losses in the equation for the atomic field, and use the \emph{total density of atoms plus molecules} in the loss term as the object that corresponds most logically to the measurable density of atoms.  Since the atomic field is very much larger than the molecular field in the situations we are considering, this kind of model should be a reasonably accurate approximation.

In principle, the formulation of the theory in terms of atoms and molecules provides an opportunity to give a model of three-body loss which would incorporate the actual mechanism of three-body loss as arising from inelastic collisions between atoms and molecules. In such a collision, both the atom and the molecule would normally be transferred to untrapped states, and be lost from the system---equivalent to a loss of three atoms.  We hope to develop this kind of model in a future publication.

\subsection{Simulations of structureless atoms}
For comparison with the simulations based on our formalism, we also run simulations based on a simple GPE. This corresponds to modeling the condensate using a single field $\Psi$, and by letting the interaction strength be determined solely by the scattering length. In the c-field formalism, the equation of motion for the single-component condensate is in this case given by
\begin{eqnarray}\label{eq:gpe}
\frac{\partial\Psi\ofx}{\partial t} &=&  -\frac{\hbar^2\nabla^2}{2m}\Psi\ofx  + \mathcal{P}_a\left\{V_a\ofx\Psi\ofx + U_{0}|\Psi\ofx|^2\Psi\ofx \right\} -\textrm{i}\gamma|\Psi\ofx|^4\Psi\ofx,
\end{eqnarray}
where the parameters are the same as in Equation (\ref{eq:dPsi_dt}), except the atom-atom interaction which is now given by
\begin{equation} 
U_{0 }= \frac{4\pi\hbar^2a_s}{m\left(1-2\Lambda a_s/\pi\right)}.
\end{equation}

\section{Results of Simulations of the Mean-Field Equations}
\label{SimMeanField}
The underlying equations of motion in the c-field formalism are the same as those of mean field theory, and quantisation is introduced by the inclusion of fluctuations in the initial state.  The inclusion of the fluctuations can cause very dramatic changes in the nature of the solutions, as was found in \cite{norrie2006b}.  In the Bragg scattering problem under study here, we have found that the effects of the quantum fluctuations are in fact rather small.  It is therefore logical to study first the solutions of the equations in the absence of the added noise in the initial conditions,
which amounts to a mean-field description of the system of atoms and molecules.
Indeed we find that these simulations provide a very satisfactory description of the problem, which agrees very well with the experimental results of \cite{papp2008}.  The effect of the noise terms is thus a matter of determining relatively small corrections to the mean field theory, and this will be done in the following section.

\subsection{General behaviour}
\label{sec:ResultsBehaviour}%
\begin{figure}[t]
\begin{center}
\subfigure[$t=0$ ms]{
\begin{psfrags}%
\psfrag{s05}[t][t]{\color[rgb]{0,0,0}\setlength{\tabcolsep}{0pt}\begin{tabular}{c}$x$ $[x_0]$\end{tabular}}%
\psfrag{s06}[b][b]{\color[rgb]{0,0,0}\setlength{\tabcolsep}{0pt}\begin{tabular}{c}$|\psi|^2 +2|\phi|^2$\end{tabular}}%
\psfrag{s07}[b][b]{\color[rgb]{0,0,0}\setlength{\tabcolsep}{0pt}\begin{tabular}{c}Axial direction\end{tabular}}%
\psfrag{s09}[t][t]{\color[rgb]{0,0,0}\setlength{\tabcolsep}{0pt}\begin{tabular}{c}$y$ $[x_0]$\end{tabular}}%
\psfrag{s10}[b][b]{\color[rgb]{0,0,0}\setlength{\tabcolsep}{0pt}\begin{tabular}{c}$|\psi|^2 +2|\phi|^2$\end{tabular}}%
\psfrag{s11}[b][b]{\color[rgb]{0,0,0}\setlength{\tabcolsep}{0pt}\begin{tabular}{c}Radial direction\end{tabular}}%
\psfrag{x01}[t][t]{0}%
\psfrag{x02}[t][t]{0.1}%
\psfrag{x03}[t][t]{0.2}%
\psfrag{x04}[t][t]{0.3}%
\psfrag{x05}[t][t]{0.4}%
\psfrag{x06}[t][t]{0.5}%
\psfrag{x07}[t][t]{0.6}%
\psfrag{x08}[t][t]{0.7}%
\psfrag{x09}[t][t]{0.8}%
\psfrag{x10}[t][t]{0.9}%
\psfrag{x11}[t][t]{1}%
\psfrag{x12}[t][t]{-5}%
\psfrag{x13}[t][t]{0}%
\psfrag{x14}[t][t]{5}%
\psfrag{x15}[t][t]{-100}%
\psfrag{x16}[t][t]{0}%
\psfrag{x17}[t][t]{100}%
\psfrag{v01}[r][r]{0}%
\psfrag{v02}[r][r]{0.2}%
\psfrag{v03}[r][r]{0.4}%
\psfrag{v04}[r][r]{0.6}%
\psfrag{v05}[r][r]{0.8}%
\psfrag{v06}[r][r]{1}%
\psfrag{v07}[r][r]{0}%
\psfrag{v08}[r][r]{20}%
\psfrag{v09}[r][r]{40}%
\psfrag{v10}[r][r]{60}%
\psfrag{v11}[r][r]{0}%
\psfrag{v12}[r][r]{20}%
\psfrag{v13}[r][r]{40}%
\psfrag{v14}[r][r]{60}%
\resizebox{12cm}{!}{\includegraphics{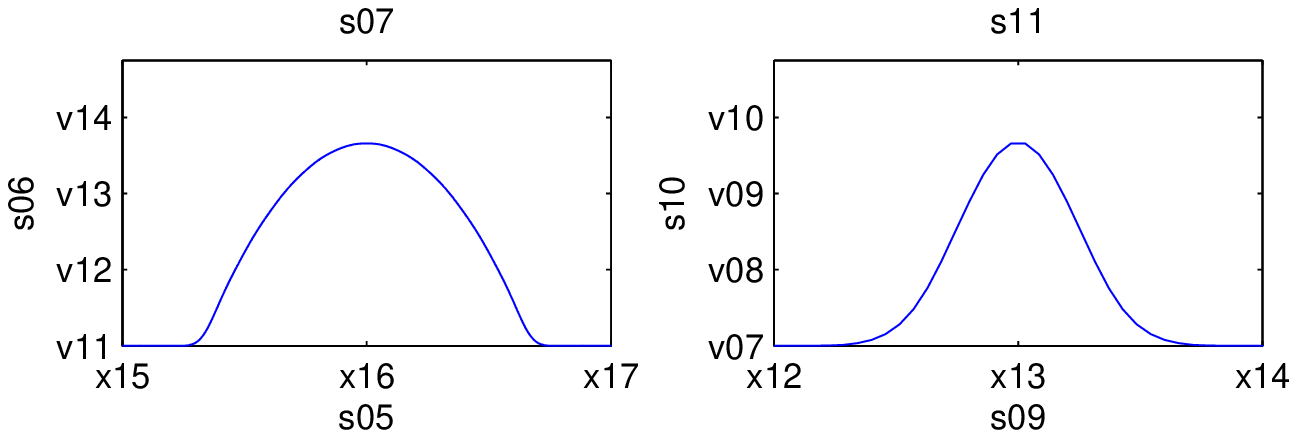}}%
\end{psfrags}%
\label{fig:profile0}}\vspace{0.4in}
\subfigure[$t=0.09$ ms]{
\begin{psfrags}%
\psfrag{s05}[t][t]{\color[rgb]{0,0,0}\setlength{\tabcolsep}{0pt}\begin{tabular}{c}$x$ $[x_0]$\end{tabular}}%
\psfrag{s06}[b][b]{\color[rgb]{0,0,0}\setlength{\tabcolsep}{0pt}\begin{tabular}{c}$|\psi|^2 +2|\phi|^2$\end{tabular}}%
\psfrag{s07}[b][b]{\color[rgb]{0,0,0}\setlength{\tabcolsep}{0pt}\begin{tabular}{c}Axial direction\end{tabular}}%
\psfrag{s09}[t][t]{\color[rgb]{0,0,0}\setlength{\tabcolsep}{0pt}\begin{tabular}{c}$y$ $[x_0]$\end{tabular}}%
\psfrag{s10}[b][b]{\color[rgb]{0,0,0}\setlength{\tabcolsep}{0pt}\begin{tabular}{c}$|\psi|^2 +2|\phi|^2$\end{tabular}}%
\psfrag{s11}[b][b]{\color[rgb]{0,0,0}\setlength{\tabcolsep}{0pt}\begin{tabular}{c}Radial direction\end{tabular}}%
\psfrag{x01}[t][t]{0}%
\psfrag{x02}[t][t]{0.1}%
\psfrag{x03}[t][t]{0.2}%
\psfrag{x04}[t][t]{0.3}%
\psfrag{x05}[t][t]{0.4}%
\psfrag{x06}[t][t]{0.5}%
\psfrag{x07}[t][t]{0.6}%
\psfrag{x08}[t][t]{0.7}%
\psfrag{x09}[t][t]{0.8}%
\psfrag{x10}[t][t]{0.9}%
\psfrag{x11}[t][t]{1}%
\psfrag{x12}[t][t]{-5}%
\psfrag{x13}[t][t]{0}%
\psfrag{x14}[t][t]{5}%
\psfrag{x15}[t][t]{-100}%
\psfrag{x16}[t][t]{0}%
\psfrag{x17}[t][t]{100}%
\psfrag{v01}[r][r]{0}%
\psfrag{v02}[r][r]{0.2}%
\psfrag{v03}[r][r]{0.4}%
\psfrag{v04}[r][r]{0.6}%
\psfrag{v05}[r][r]{0.8}%
\psfrag{v06}[r][r]{1}%
\psfrag{v07}[r][r]{0}%
\psfrag{v08}[r][r]{20}%
\psfrag{v09}[r][r]{40}%
\psfrag{v10}[r][r]{60}%
\psfrag{v11}[r][r]{0}%
\psfrag{v12}[r][r]{20}%
\psfrag{v13}[r][r]{40}%
\psfrag{v14}[r][r]{60}%
\resizebox{12cm}{!}{\includegraphics{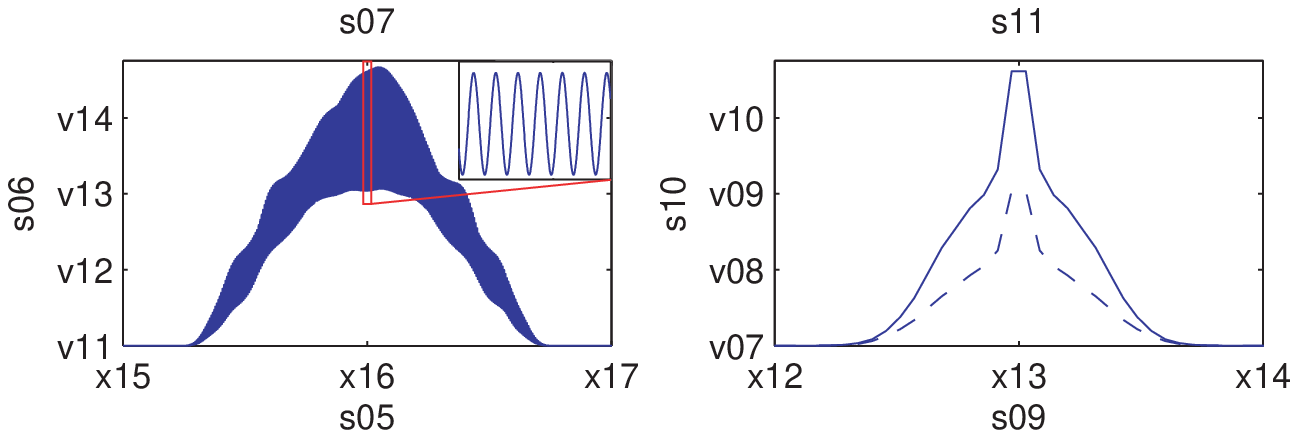}}%
\end{psfrags}%
\label{fig:profile076}}
\caption{Coordinate space profiles for $a_s = 890a_0$ at a time (a) just before the Bragg pulse is applied, and (b) at the end of the Bragg pulse. The left panels show a slice of the condensate in the axial direction, i.e. the direction of the pulse, and the right panels show slices in the radial direction. In the bottom right panel we show two slices, corresponding to the crest (solid line) and trough (dashed line) of the interference fringe centered around $x = 0$. The inset in the bottom left panel shows the profile appearance in the region marked by the red box. The parameter $x_0$ is the length scale associated with the $x$-axis of the trap, given by $x_0 = \sqrt{\hbar/2m\omega_x}\approx6.665\times10^{-7}$m.}
\label{fig:Profiles}
\end{center}
\end{figure}

The coordinate space profiles from a typical simulation run are shown in Fig.\,\ref{fig:Profiles}. The scattering length is in this case $890a_0$. Fig.\,\ref{fig:profile0} shows the radial and axial profiles of the condensate after it has been ramped to the scattering length of interest, at the moment just before the Bragg pulse is applied. Fig.\,\ref{fig:profile076} shows the the profiles for the same simulation run, at the end of the Bragg pulse.

As can be clearly seen in Fig.\,\ref{fig:profile0}, the condensate profile in the axial direction before the onset of the pulse is similar in shape to a Thomas-Fermi profile, whereas in the radial direction it is more like a Gaussian. This is a result of the elongated shape of the condensate, due to the aspect ratio of the trap.
As noted in Sect.\ref{Interpreting the Experiment}, this makes any estimate of the average density difficult to justify.

The Bragg pulse is applied in the axial direction; in Fig.\,\ref{fig:profile076} we can clearly see the effect of this as interference fringes in the axial profile of the condensate. In the radial direction we have therefore plotted two distinctly different profiles, corresponding to the crest and trough of the central fringe. We obtain similar profiles with large density variations for each of the different scattering lengths in our simulations.

The particle losses arising from three-body recombination events and the change in density in our simulations are very different 
from those predicted by \cite{papp2008}, whose prediction is that the density will change by ``less than 30\%''.  In contrast, in our simulations the density changes by up to 70\% of the initial density (see Tab.\,\ref{tab:change}.) Furthermore, in the experiment, the three-body loss is observed to be ``typically <30\%'' \cite{papp2008}; whereas in our simulations the losses never 
exceed 10\% of the total atom number. We believe that the main reason for these differences is the inappropriate model used to describe the condensate in \cite{papp2008}. However, there is also a significant degree of uncertainty in our calculations of the three-body loss, because of the lack of accurate data for the loss rate, and this could also contribute to the discrepancy.

\begin{table}[t]
\caption{\label{tab:change}Condensate three-body loss and density change during the Bragg pulse, for the different scattering lengths.}
\begin{center}\small
\begin{tabular}{c|l|c}\hline
Scattering length $a_s$ [$a_0$]&Three-body loss&Density change \\
\hline
150&\hspace{3em}1\%&70\% \\
300&\hspace{3em}2.5\%& 50\%\\
500&\hspace{3em}2.5\%& 33\%\\
585&\hspace{3em}4\%& 30\%\\
695&\hspace{3em}5\%& 30\%\\
805&\hspace{3em}5.5\%& 26\%\\
890&\hspace{3em}6\%& 24\%\\
\hline
\end{tabular}
\end{center}
\end{table}
\begin{figure}[b]
\begin{center}
\begin{psfrags}%
\psfrag{s01}[t][t]{\color[rgb]{0,0,0}\setlength{\tabcolsep}{0pt}\begin{tabular}{c}Bragg frequency [kHz]\end{tabular}}%
\psfrag{s02}[b][b]{\color[rgb]{0,0,0}\setlength{\tabcolsep}{0pt}\begin{tabular}{c}Momentum transfer $[N|\q|]$\end{tabular}}%
\psfrag{x01}[t][t]{0}%
\psfrag{x02}[t][t]{0.1}%
\psfrag{x03}[t][t]{0.2}%
\psfrag{x04}[t][t]{0.3}%
\psfrag{x05}[t][t]{0.4}%
\psfrag{x06}[t][t]{0.5}%
\psfrag{x07}[t][t]{0.6}%
\psfrag{x08}[t][t]{0.7}%
\psfrag{x09}[t][t]{0.8}%
\psfrag{x10}[t][t]{0.9}%
\psfrag{x11}[t][t]{1}%
\psfrag{x12}[t][t]{5}%
\psfrag{x13}[t][t]{10}%
\psfrag{x14}[t][t]{15}%
\psfrag{x15}[t][t]{20}%
\psfrag{x16}[t][t]{25}%
\psfrag{x17}[t][t]{30}%
\psfrag{v01}[r][r]{0}%
\psfrag{v02}[r][r]{0.1}%
\psfrag{v03}[r][r]{0.2}%
\psfrag{v04}[r][r]{0.3}%
\psfrag{v05}[r][r]{0.4}%
\psfrag{v06}[r][r]{0.5}%
\psfrag{v07}[r][r]{0.6}%
\psfrag{v08}[r][r]{0.7}%
\psfrag{v09}[r][r]{0.8}%
\psfrag{v10}[r][r]{0.9}%
\psfrag{v11}[r][r]{1}%
\psfrag{v12}[r][r]{0}%
\psfrag{v13}[r][r]{0.01}%
\psfrag{v14}[r][r]{0.02}%
\psfrag{v15}[r][r]{0.03}%
\psfrag{v16}[r][r]{0.04}%
\psfrag{v17}[r][r]{0.05}%
\psfrag{v18}[r][r]{0.06}%
\psfrag{v19}[r][r]{0.07}%
\resizebox{10cm}{!}{\includegraphics{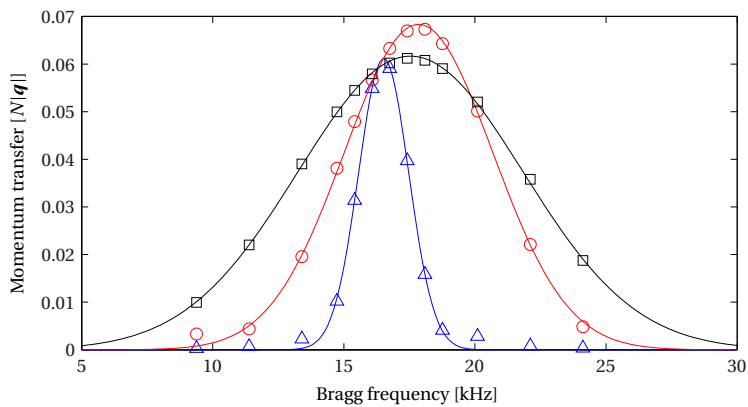}}%
\end{psfrags}%
\caption{The Bragg spectra for three different values of the scattering length: 150$a_0$ (blue triangles), 500$a_0$ (red circles) and 890$a_0$ (black squares). The solid lines are Gaussian fits to the data points. }
\label{fig:spectra}
\end{center}
\end{figure}
\subsection{Bragg spectra and lineshift}
The Bragg spectrum is obtained by changing the frequency difference $\omega$ in (\ref{eq:braggpulse}), and calculating the momentum transferred to the condensate for each frequency. We calculate the normalized momentum transfer as
\begin{equation}\label{momentum-transfer}
P(t)=\frac{1}{N(t)|\q|}\int{d\kbf\left(|\psi(\kbf,t)|^2+2|\phi(\kbf,t)|^2\right)\kbf}.
\end{equation}

Typical spectra from our simulations are shown in Fig.\,\ref{fig:spectra}. The difference in width between the different spectra is due to the change in duration of the Bragg pulse. The peaks of the Bragg spectra in Fig.\,\ref{fig:spectra} are shifted from the position of the corresponding peaks for the non-interacting gas, located at approximately 15.4 kHz.

\begin{figure}[t]
\begin{center}
\begin{psfrags}%
\psfragscanon%
\psfrag{s05}[t][t]{\color[rgb]{0,0,0}\setlength{\tabcolsep}{0pt}\begin{tabular}{c}Scattering length $a_s$ [$a_0$]\end{tabular}}%
\psfrag{s06}[b][b]{\color[rgb]{0,0,0}\setlength{\tabcolsep}{0pt}\begin{tabular}{c}Line shift [kHz]\end{tabular}}%
\psfrag{s10}[][]{\color[rgb]{0,0,0}\setlength{\tabcolsep}{0pt}\begin{tabular}{c} \end{tabular}}%
\psfrag{s11}[][]{\color[rgb]{0,0,0}\setlength{\tabcolsep}{0pt}\begin{tabular}{c} \end{tabular}}%
\psfrag{s12}[l][l]{\color[rgb]{0,0,0}Atom-molecule simulation}%
\psfrag{s13}[l][l]{\color[rgb]{0,0,0}Experimental data}%
\psfrag{s14}[l][l]{\color[rgb]{0,0,0}Simple shift prediction}%
\psfrag{s15}[l][l]{\color[rgb]{0,0,0}Atom Bogoliubov}%
\psfrag{s16}[l][l]{\color[rgb]{0,0,0}Atom-molecule Bogoliubov}%
\psfrag{s17}[l][l]{\color[rgb]{0,0,0}Atom simulation}%
\psfrag{s18}[l][l]{\color[rgb]{0,0,0}Atom-molecule simulation}%
\psfrag{x01}[t][t]{0}%
\psfrag{x02}[t][t]{0.1}%
\psfrag{x03}[t][t]{0.2}%
\psfrag{x04}[t][t]{0.3}%
\psfrag{x05}[t][t]{0.4}%
\psfrag{x06}[t][t]{0.5}%
\psfrag{x07}[t][t]{0.6}%
\psfrag{x08}[t][t]{0.7}%
\psfrag{x09}[t][t]{0.8}%
\psfrag{x10}[t][t]{0.9}%
\psfrag{x11}[t][t]{1}%
\psfrag{x12}[t][t]{0}%
\psfrag{x13}[t][t]{100}%
\psfrag{x14}[t][t]{200}%
\psfrag{x15}[t][t]{300}%
\psfrag{x16}[t][t]{400}%
\psfrag{x17}[t][t]{500}%
\psfrag{x18}[t][t]{600}%
\psfrag{x19}[t][t]{700}%
\psfrag{x20}[t][t]{800}%
\psfrag{x21}[t][t]{900}%
\psfrag{v01}[r][r]{0}%
\psfrag{v02}[r][r]{0.1}%
\psfrag{v03}[r][r]{0.2}%
\psfrag{v04}[r][r]{0.3}%
\psfrag{v05}[r][r]{0.4}%
\psfrag{v06}[r][r]{0.5}%
\psfrag{v07}[r][r]{0.6}%
\psfrag{v08}[r][r]{0.7}%
\psfrag{v09}[r][r]{0.8}%
\psfrag{v10}[r][r]{0.9}%
\psfrag{v11}[r][r]{1}%
\psfrag{v12}[r][r]{0}%
\psfrag{v13}[r][r]{1}%
\psfrag{v14}[r][r]{2}%
\psfrag{v15}[r][r]{3}%
\psfrag{v16}[r][r]{4}%
\psfrag{v17}[r][r]{5}%
\psfrag{v18}[r][r]{6}%
\psfrag{v19}[r][r]{7}%
\resizebox{12cm}{!}{\includegraphics{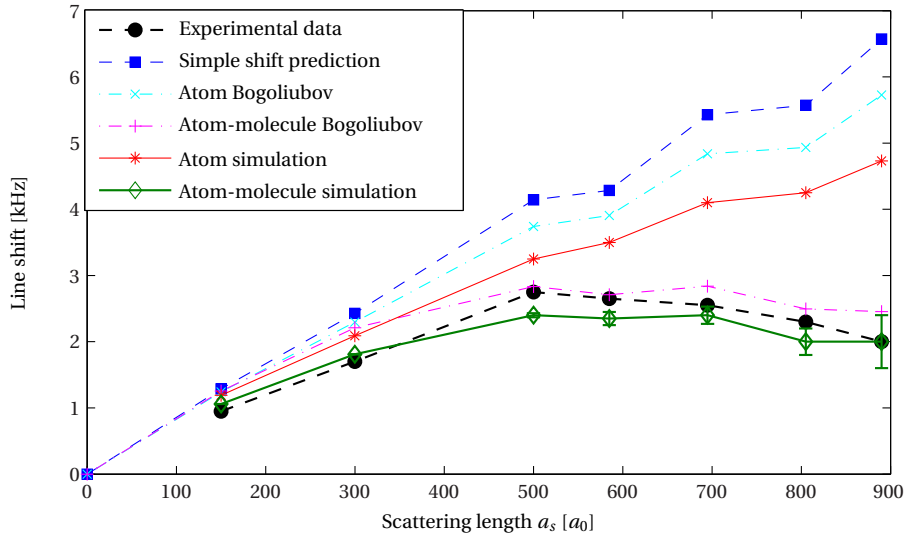}}%
\end{psfrags}%
\caption{The shift of the peak of the Bragg spectra for different values of the scattering length. The solid lines are results from our simulations, the dashed lines are the experimental results and predictions presented in \cite{papp2008} and the dash-dotted lines are calculations based on the Bogoliubov treatment of \paperii. Our atom-molecule simulation (green diamonds) is significantly different from that of the structureless atom (red stars), but agrees well with both the experimental data (black circles) and our atom-molecule Bogoliubov calculation (magenta plusses). The prediction of the lineshift based on the excitation spectrum in the large $k$ limit of equation (\ref{eq:simpleshift}) (blue squares) shows a very different behaviour for large scattering lengths, as does the structureless Bogoliubov calculation (cyan crosses). The error bars on our atom-molecule calculation indicate the uncertainty in the experimental estimates of the density, as is shown in Tab.\,\ref{tab:density}; similar error estimates would apply to the other curves, but have been omitted for clarity.}
\label{fig:lineshift}
\end{center}
\end{figure}

In Fig.\,\ref{fig:lineshift} the shifts of the Bragg spectra from result for the noninteracting case are plotted as a function of the scattering length. 

For comparison, we have also plotted the experimental results of the Bragg lineshift from Ref. \cite{papp2008}. Fig.\,\ref{fig:lineshift} also includes the theoretical prediction of the lineshift based on (\ref{eq:simpleshift}), but using the density weighted density instead of the space-averaged density used in the experimental paper. For a Thomas--Fermi profile the density weighted density is $10/7$ times larger than the space-averaged density, as discussed in Sect.\,\ref{sec:SimInitial}. This correction eliminates the anomaly apparent in Fig.\,3(a) in \cite{papp2008}, in which the experimental data  and the simple shift prediction agree almost perfectly up to a scattering length of about $500a_0$, and then deviate sharply.  Using the density-averaged density, a smooth increase in deviation is apparent.

As can be seen clearly in the figure, the simulations based on our model show quantitative agreement with the experimental data. We have included error bars on the experimental data points; these indicate the uncertainty in the experimental estimates of the density in the experiment. Similar error bars should therefore also be included on the other lines in Fig.\,\ref{fig:lineshift}, but we have omitted these for clarity. 

For small scattering lengths, the molecule field is very small, but neverthless plays an important role since its presence gives rise to a positive scattering length, in contrast to the negative background scattering length. The fact that the binding energy is larger at low scattering lengths makes it possible for the molecule field to adiabatically follow the atom field, and thus the condensate behaviour is very similar to that predicted by a GPE description. This is clear in Fig.\,\ref{fig:lineshift}, where we have included the result from the simulations based on the GPE (\ref{eq:gpe}). At larger scattering lengths, the bound state evolves more slowly and the atom-molecule simulations become very different from those for structureless atoms. 

Finally, in Fig.\,\ref{fig:lineshift}, we have included the results from Bogoliubov treatments of the ideal case of a uniform condensate, both for the case of a single atom field, as in \cite{blakie2000,blakie2002}, and in the case of an atom-molecule system, as in \paperii. We find that the atom-molecule Bogoliubov treatment shows surprisingly close agreement with both our simulations and with the experimental data.

\section{Results of Full C-Field Simulations}
\label{sec:Noise}
In the c-field methods, the effect of quantum fluctuations is included by adding stochastic terms to the initial state, corresponding to on average half a particle per mode. 
In our treatment of Bragg scattering, we will follow the approximate procedure as noted in 
\cite{blakie2008} of adding Gaussian random noise, $r\ofk$ and $s\ofk$, with zero mean and unit standard deviation  to the initial momentum amplitudes for the atoms $\psi_0$ and $\phi_0$, according to
\begin{eqnarray}
\psi\ofk = \psi_0\ofk + \frac{{r}\ofk}{\sqrt{2}}, \qquad
\phi\ofk  =  \phi_0\ofk + \frac{{s}\ofk}{\sqrt{2}}.
\end{eqnarray}
Each simulation run can then be seen as corresponding to a single run of an experiment, and the expectation values of observables are obtained by taking the average of several different runs. The average of the noise amplitudes is obviously
\begin{equation}
\left\langle \left|\frac{{r}\ofk}{\sqrt{2}}\right|^2\right\rangle = \left\langle \left|\frac{{s}\ofk}{\sqrt{2}}\right|^2\right\rangle = \frac{1}{2},
\end{equation}
corresponding to half a noise particle per mode.

\begin{figure}[t]
\begin{center}
\begin{psfrags}%
\psfrag{s07}[t][t]{\color[rgb]{0,0,0}\setlength{\tabcolsep}{0pt}\begin{tabular}{c}$x$ $[x_0]$\end{tabular}}%
\psfrag{s08}[b][b]{\color[rgb]{0,0,0}\setlength{\tabcolsep}{0pt}\begin{tabular}{c}$y$ $[x_0]$\end{tabular}}%
\psfrag{s09}[b][b]{\color[rgb]{0,0,0}\setlength{\tabcolsep}{0pt}\begin{tabular}{c}arg$(\psi)$\end{tabular}}%
\psfrag{s12}[][]{\color[rgb]{0,0,0}\setlength{\tabcolsep}{0pt}\begin{tabular}{c} \end{tabular}}%
\psfrag{s13}[][]{\color[rgb]{0,0,0}\setlength{\tabcolsep}{0pt}\begin{tabular}{c} \end{tabular}}%
\psfrag{s14}[t][t]{\color[rgb]{0,0,0}\setlength{\tabcolsep}{0pt}\begin{tabular}{c}$x$ $[x_0]$\end{tabular}}%
\psfrag{s15}[b][b]{\color[rgb]{0,0,0}\setlength{\tabcolsep}{0pt}\begin{tabular}{c}$y$ $[x_0]$\end{tabular}}%
\psfrag{s16}[b][b]{\color[rgb]{0,0,0}\setlength{\tabcolsep}{0pt}\begin{tabular}{c}arg$(\phi)$\end{tabular}}%
\psfrag{s19}[][]{\color[rgb]{0,0,0}\setlength{\tabcolsep}{0pt}\begin{tabular}{c} \end{tabular}}%
\psfrag{s20}[][]{\color[rgb]{0,0,0}\setlength{\tabcolsep}{0pt}\begin{tabular}{c} \end{tabular}}%
\psfrag{x01}[t][t]{0}%
\psfrag{x02}[t][t]{0.1}%
\psfrag{x03}[t][t]{0.2}%
\psfrag{x04}[t][t]{0.3}%
\psfrag{x05}[t][t]{0.4}%
\psfrag{x06}[t][t]{0.5}%
\psfrag{x07}[t][t]{0.6}%
\psfrag{x08}[t][t]{0.7}%
\psfrag{x09}[t][t]{0.8}%
\psfrag{x10}[t][t]{0.9}%
\psfrag{x11}[t][t]{1}%
\psfrag{x12}[t][t]{0}%
\psfrag{x13}[t][t]{0.5}%
\psfrag{x14}[t][t]{1}%
\psfrag{x15}[t][t]{0}%
\psfrag{x16}[t][t]{0.5}%
\psfrag{x17}[t][t]{1}%
\psfrag{x18}[t][t]{-60}%
\psfrag{x19}[t][t]{-40}%
\psfrag{x20}[t][t]{-20}%
\psfrag{x21}[t][t]{0}%
\psfrag{x22}[t][t]{20}%
\psfrag{x23}[t][t]{40}%
\psfrag{x24}[t][t]{60}%
\psfrag{x25}[t][t]{-60}%
\psfrag{x26}[t][t]{-40}%
\psfrag{x27}[t][t]{-20}%
\psfrag{x28}[t][t]{0}%
\psfrag{x29}[t][t]{20}%
\psfrag{x30}[t][t]{40}%
\psfrag{x31}[t][t]{60}%
\psfrag{v01}[r][r]{0}%
\psfrag{v02}[r][r]{0.1}%
\psfrag{v03}[r][r]{0.2}%
\psfrag{v04}[r][r]{0.3}%
\psfrag{v05}[r][r]{0.4}%
\psfrag{v06}[r][r]{0.5}%
\psfrag{v07}[r][r]{0.6}%
\psfrag{v08}[r][r]{0.7}%
\psfrag{v09}[r][r]{0.8}%
\psfrag{v10}[r][r]{0.9}%
\psfrag{v11}[r][r]{1}%
\psfrag{v12}[l][l]{$-\pi$}%
\psfrag{v13}[l][l]{0}%
\psfrag{v14}[l][l]{$\pi$}%
\psfrag{v15}[l][l]{$-\pi$}%
\psfrag{v16}[l][l]{0}%
\psfrag{v17}[l][l]{$\pi$}%
\psfrag{v18}[r][r]{-5}%
\psfrag{v19}[r][r]{0}%
\psfrag{v20}[r][r]{5}%
\psfrag{v21}[r][r]{-5}%
\psfrag{v22}[r][r]{0}%
\psfrag{v23}[r][r]{5}%
\resizebox{12cm}{!}{\includegraphics{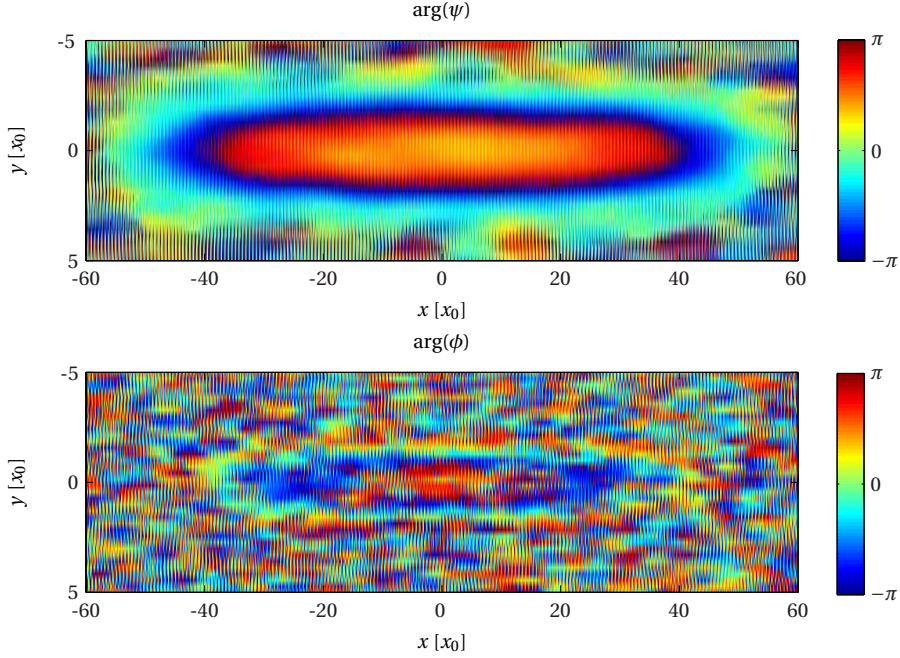}}%
\end{psfrags}%
\caption{Phase plots of a slice of the condensate centered around $z = 0$ for a scattering length of $a_s= 890a_0$, showing the atom part (top panel) and the molecule part (bottom panel) of the system. The total number of particles is around 40,000, with approximately 34,000 atoms in the form of atoms, and 6,000 atoms as molecules. The total number of noise particles is 30,000 for the atom field and 500,000 for the molecule field. The parameter $x_0$ is the length scale associated with the $x$-axis of the trap, given by $x_0 = \sqrt{\hbar/2m\omega_x}\approx6.665\times10^{-7}$m.}
\label{fig:Phase}
\end{center}
\end{figure}

Fig.\,\ref{fig:Phase} shows the phase of the spatial atom and molecule fields for a slice in the $xy$-plane for the same system as in Fig.\,\ref{fig:Profiles} at the end of the Bragg pulse. Since the initial stochastic terms are added to the \emph{momentum space} wavefunctions, and since the molecule projector encompasses a much larger part of momentum space than the atom one, there are many more noise particles in the molecule field than in the atom one. Despite this, and the fact that the molecule field is much smaller than the atom one, there is still a clearly visible phase coherence in the molecule field.

It is remarkable that the noise evident in the phase of the molecule field has very little effect on the results of simulations.  The large positive scattering length arises directly from the population of the molecule field, and one might have expected its value to be significantly affected by the quantum fluctuations as they appear in the c-field model.

\subsection{Density Weighted Density in Terms of C-Fields}
The correct computation of the density-weighted density involves some care, since it
involves products of four field operators, including both molecule and field operators.  The details of how this is done are presented in Appendix~\ref{sec:AppDensity}, whose results are in summary:
\begin{enumerate}
\item The average total particle number in the noise simulations is given by
\begin{equation}
\overline{{N}(t)} = \left\langle  \int{d\x\left[\left\{\hat{n}_a(t)\right\}_{\text{sym}} + 2\left\{\hat{n}_m(t)\right\}_{\text{sym}} - \frac{\Delta_a}{2} - \Delta_m \right]} \right\rangle,
\end{equation}
where $\left\{\hat{n}_a(\x,t)\right\}_{\text{sym}}$ and $\left\{\hat{n}_m(\x,t)\right\}_{\text{sym}}$ are the symmetrically ordered averages
\begin{eqnarray}
\left\{\hat{n}_a(\x,t)\right\}_{\text{sym}} &\equiv& \left\{\hat\psi^\dagger(\x,t)\hat\psi(\x,t)\right\}_{\text{sym}}, \\
\left\{\hat{n}_m(\x,t)\right\}_{\text{sym}} &\equiv& \left\{\hat\phi^\dagger(\x,t)\hat\phi(\x,t)\right\}_{\text{sym}}. 
\end{eqnarray}
The parameters $\Delta_a$ and $\Delta_m$ corresponds to the noise on the atom and molecule coordinate space wavefunctions, respectively, given by
\begin{eqnarray}
\left\langle \left|r\ofk\right|^2\right\rangle &=& \Delta_a, \\
\left\langle \left|s\ofk\right|^2\right\rangle &=& \Delta_m.
\end{eqnarray}
Since the molecule field has much more initial noise added to it, $\Delta_m$ is much larger than $\Delta_a$. 

\item  The coordinate space density-weighted density is given by
\begin{eqnarray}
\label{eq:DWdensityTW}
\overline{{n}(t)} &=& \frac{1}{\overline{{N}(t)}} \left\langle \int d\x\left[\overline{{n}^2_a(t)}+ 4\overline{{n}^2_m(t)}- 2\Delta_m\left\{\hat{n}_a(t)\right\}_{\text{sym}} -  2\Delta_a\left\{\hat{n}_m(t)\right\}_{\text{sym}} \right. \right. \nonumber \\
&& \left.\left. + 4 \left\{\hat{n}_a(t)\right\}_{\text{sym}} \left\{\hat{n}_m(t)\right\}_{\text{sym}}  +\Delta_a\Delta_m \right] \right\rangle,
\end{eqnarray}
where $\overline{n^2_a(\x,t)}$ and $\overline{n^2_m(\x,t)}$ are given by
\begin{eqnarray}
\overline{n^2_a(\x,t)} &=& \left\{\hat{n}^2_a(\x,t)\right\}_{\text{sym}} -2\Delta_a\left\{\hat{n}_a(\x,t)\right\}_{\text{sym}}+\frac{\Delta_a^2}{2}, \\
\overline{n^2_m(\x,t)} &=& \left\{\hat{n}^2_m(\x,t)\right\}_{\text{sym}} -2\Delta_m\left\{\hat{n}_m(\x,t)\right\}_{\text{sym}}+\frac{\Delta_m^2}{2}, 
\end{eqnarray}
where 
\begin{eqnarray}
\left\{\hat{n}^2_a(\x,t)\right\}_{\text{sym}} &\equiv& \left\{\hat\psi^{\dagger2}(\x,t)\hat\psi^2(\x,t)\right\}_{\text{sym}}, \\
\left\{\hat{n}^2_m(\x,t)\right\}_{\text{sym}} &\equiv& \left\{\hat\phi^{\dagger2}(\x,t)\hat\phi^2(\x,t)\right\}_{\text{sym}}. 
\end{eqnarray}
\end{enumerate}

\begin{figure}[t]
\begin{center}
\begin{psfrags}%
\psfrag{s05}[t][t]{\color[rgb]{0,0,0}\setlength{\tabcolsep}{0pt}\begin{tabular}{c}t [ms]\end{tabular}}%
\psfrag{s06}[b][b]{\color[rgb]{0,0,0}\setlength{\tabcolsep}{0pt}\begin{tabular}{c}density $[10^{19}m^-3]$\end{tabular}}%
\psfrag{s10}[][]{\color[rgb]{0,0,0}\setlength{\tabcolsep}{0pt}\begin{tabular}{c} \end{tabular}}%
\psfrag{s11}[][]{\color[rgb]{0,0,0}\setlength{\tabcolsep}{0pt}\begin{tabular}{c} \end{tabular}}%
\psfrag{s12}[l][l]{\color[rgb]{0,0,0}Noise time-average}%
\psfrag{s13}[l][l]{\color[rgb]{0,0,0}Noise-free simulation}%
\psfrag{s14}[l][l]{\color[rgb]{0,0,0}Noise-free time-average}%
\psfrag{s15}[l][l]{\color[rgb]{0,0,0}Noise simulations}%
\psfrag{s16}[l][l]{\color[rgb]{0,0,0}Noise time-average}%
\psfrag{x01}[t][t]{0}%
\psfrag{x02}[t][t]{0.1}%
\psfrag{x03}[t][t]{0.2}%
\psfrag{x04}[t][t]{0.3}%
\psfrag{x05}[t][t]{0.4}%
\psfrag{x06}[t][t]{0.5}%
\psfrag{x07}[t][t]{0.6}%
\psfrag{x08}[t][t]{0.7}%
\psfrag{x09}[t][t]{0.8}%
\psfrag{x10}[t][t]{0.9}%
\psfrag{x11}[t][t]{1}%
\psfrag{x12}[t][t]{0}%
\psfrag{x13}[t][t]{0.02}%
\psfrag{x14}[t][t]{0.04}%
\psfrag{x15}[t][t]{0.06}%
\psfrag{x16}[t][t]{0.08}%
\psfrag{v01}[r][r]{0}%
\psfrag{v02}[r][r]{0.1}%
\psfrag{v03}[r][r]{0.2}%
\psfrag{v04}[r][r]{0.3}%
\psfrag{v05}[r][r]{0.4}%
\psfrag{v06}[r][r]{0.5}%
\psfrag{v07}[r][r]{0.6}%
\psfrag{v08}[r][r]{0.7}%
\psfrag{v09}[r][r]{0.8}%
\psfrag{v10}[r][r]{0.9}%
\psfrag{v11}[r][r]{1}%
\psfrag{v12}[r][r]{7.5}%
\psfrag{v13}[r][r]{8}%
\psfrag{v14}[r][r]{8.5}%
\psfrag{v15}[r][r]{9}%
\psfrag{v16}[r][r]{9.5}%
\psfrag{v17}[r][r]{10}%
\psfrag{v18}[r][r]{10.5}%
\psfrag{v19}[r][r]{11}%
\resizebox{12cm}{!}{\includegraphics{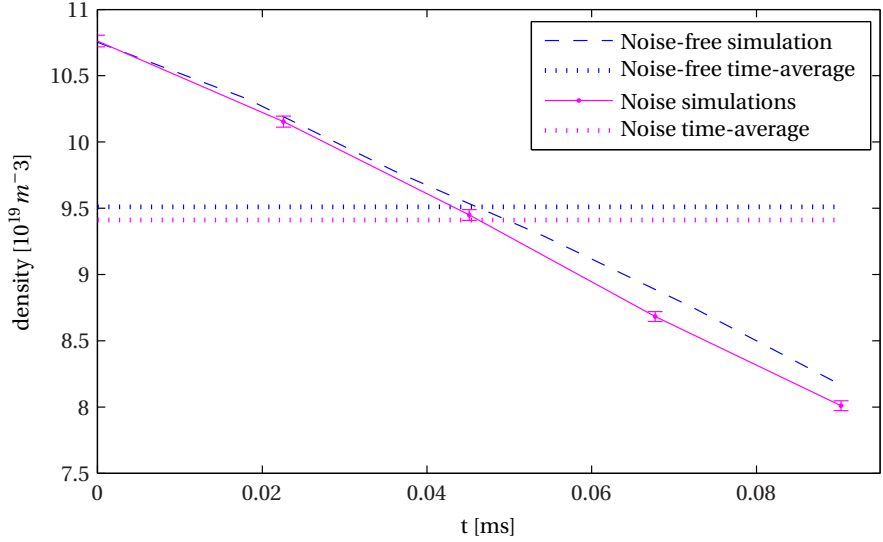}}%
\end{psfrags}%
\caption{Time evolution of the density-weighted density during the application of the Bragg pulse to a condensate. The scattering length is $a_s = 890a_0$, and the pulse length is $0.09\text{ms}$. The solid line is calculated using (\ref{eq:DWdensityTW}) for 30 different runs, with error bars indicating the statistical error from these runs. The dashed line shows the density-weighted density from a simulation without initial noise. The dotted lines shows the time-average of the density-weighted density for the noise free simulation (blue) and for the full simulations (magenta).}
\label{fig:Density}
\end{center}
\end{figure}

We run 30 simulations with noise and compare the density-weighted density obtained from these runs using (\ref{eq:DWdensityTW}) with that obtained in a single simulation run without any stochastic terms added to the initial state. The result for one of these comparisons is shown in Fig.\,\ref{fig:Density}, where we have plotted the evolution of the density-weighted density for a scattering length of $a_s = 890a_0$. 
As can be seen clearly in the figure, the initial density-weighted density is the same for both the noise-free simulation and the average of the 30 runs with noise. However, as the condensate evolves over time, the result from the noise simulations is slightly lower than that from the noise-free run. The resulting time-average of the density-weighted density will therefore be slightly higher if we neglect the initial fluctuations, although the size of the change is much less than the experimental uncertainty.

\subsection{Bragg Spectra from C-Field Simulations}
\begin{figure}[t]
\begin{center}
\begin{psfrags}%
\psfrag{s01}[t][t]{\color[rgb]{0,0,0}\setlength{\tabcolsep}{0pt}\begin{tabular}{c}Bragg frequency [kHz]\end{tabular}}%
\psfrag{s02}[b][b]{\color[rgb]{0,0,0}\setlength{\tabcolsep}{0pt}\begin{tabular}{c}Momentum transfer\end{tabular}}%
\psfrag{x01}[t][t]{0}%
\psfrag{x02}[t][t]{0.1}%
\psfrag{x03}[t][t]{0.2}%
\psfrag{x04}[t][t]{0.3}%
\psfrag{x05}[t][t]{0.4}%
\psfrag{x06}[t][t]{0.5}%
\psfrag{x07}[t][t]{0.6}%
\psfrag{x08}[t][t]{0.7}%
\psfrag{x09}[t][t]{0.8}%
\psfrag{x10}[t][t]{0.9}%
\psfrag{x11}[t][t]{1}%
\psfrag{x12}[t][t]{10}%
\psfrag{x13}[t][t]{15}%
\psfrag{x14}[t][t]{20}%
\psfrag{x15}[t][t]{25}%
\psfrag{v01}[r][r]{0}%
\psfrag{v02}[r][r]{0.1}%
\psfrag{v03}[r][r]{0.2}%
\psfrag{v04}[r][r]{0.3}%
\psfrag{v05}[r][r]{0.4}%
\psfrag{v06}[r][r]{0.5}%
\psfrag{v07}[r][r]{0.6}%
\psfrag{v08}[r][r]{0.7}%
\psfrag{v09}[r][r]{0.8}%
\psfrag{v10}[r][r]{0.9}%
\psfrag{v11}[r][r]{1}%
\psfrag{v12}[r][r]{0}%
\psfrag{v13}[r][r]{0.01}%
\psfrag{v14}[r][r]{0.02}%
\psfrag{v15}[r][r]{0.03}%
\psfrag{v16}[r][r]{0.04}%
\psfrag{v17}[r][r]{0.05}%
\psfrag{v18}[r][r]{0.06}%
\psfrag{v19}[r][r]{0.07}%
\resizebox{12cm}{!}{\includegraphics{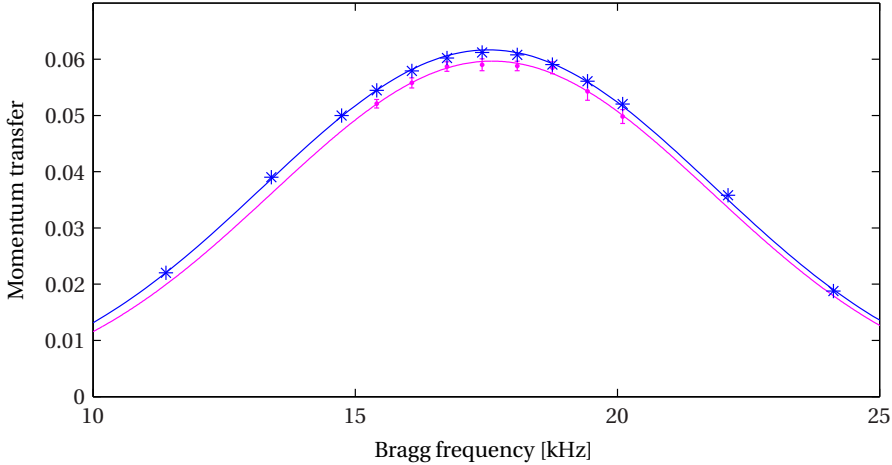}}%
\end{psfrags}%
\caption{Bragg spectrum for a scattering length of 890$a_0$, showing the result obtained from averaging over 30 runs of simulations performed with stochastic terms added to the initial modes (magenta dots), as well as the result from a single simulation without initial noise (blue stars). The solid lines are Gaussian fits to the data points.}
\label{fig:NoiseSpectra}
\end{center}
\end{figure}

For the full c-field simulations, instead of the results in 
(\ref{atomTotalNumber}, \ref{momentum-transfer}), we calculate the momentum transfer as
\begin{equation}
\label{eq:NoiseMtmTrans}
P(t)=\frac{1}{\overline{{N}(t)}|\q|}\left\langle\int{d\kbf\left(|\psi(\kbf,t)|^2+2|\phi(\kbf,t)|^2 - \frac{3}{2}\right)\kbf}\right\rangle,
\end{equation}
where the factor of $\frac{3}{2}$ is subtracted to account for the initial noise. Similarly, the total number of particles is given by
\begin{equation}
\overline{{N}(t)} = \left\langle\int{d\kbf\left(|\psi(\kbf,t)|^2+2|\phi(\kbf,t)|^2 - \frac{3}{2}\right)}\right\rangle.
\end{equation}
Fig.\,\ref{fig:NoiseSpectra} shows the momentum transfer calculated using equation (\ref{eq:NoiseMtmTrans}), where the average has been taken over 30 simulation runs. In comparison, we have also plotted the momentum transfer from a single simulation run without any initial noise terms. The noise simulations give a spectrum that is slightly narrower than the noise-free simulation and with a slightly lower amplitude. However, the position of the spectral peak is essentially the same for both the simulations with and without initial noise.

In our simulations, we find that averaging over several different noise simulations in this way gives us very results similar to those obtained by running the same simulation without including the noise. Although the vacuum fluctuations seem to have some small effect on the evolution density, overall, the effect on the condensate dynamics appears to be unimportant to the Bragg scattering experiment.  We can therefore be confident that the simulations of the mean-field equations of the atom-molecule system which we did in Sect.\,\ref{SimMeanField}---equivalent to omitting the initial quantum fluctuations---provide a reliable description of the Bragg scattering experiment.

\section{Conclusions}
The aim of the experiment of Ref. \cite{papp2008} was to to probe the behaviour of a Bose-Einstein condensate in the regime where two major simplifications normally made in its theoretical description were not valid.  These simplifications are made in terms of three dimensionless parameters:
\begin{enumerate}
\item \emph{Weak Interactions }: This requires $\sqrt{8\pi n a_s^3} \ll 1$. It is important to note that this approximation is necessary not only for the validity the Gross--Pitaevskii equation, but also for the validity of the local quantum field theory, to which the  Gross--Pitaevskii is an approximation.

In the experiment the condensate was compressed, and the scattering length increased by using a Feshbach resonance, in order to ensure the violation of this condition. 

\item \emph{Local Interactions }: By this, it is meant that the length scale on which processes of interest take place is much larger than that of the interactions.  In the experiment the momentum transfer involved in the Bragg scattering was chosen to be sufficiently large that the momentum dependence of the scattering amplitude would be important.
\end{enumerate}
In addition, the parameters of the experiment were chosen so that the the relevant quasiparticles, that is, those with momentum corresponding to the Bragg wavenumber, were definitely not in the free particle regime.

In the three papers in this series we have shown how to take account of all of these within a tractable formalism.  The most significant aspect of both the experiment and the theory is the clear demonstration that the large scattering lengths generated by Feshbach resonances do not give rise to interactions of the hard-sphere kind, as treated originally by Huang and Yang \cite{Huang1957a,Huang1957b}.  Indeed, it is remarkable that the classical Huang--Yang theory works so well for systems with Feshbach resonance enhanced interactions. For this reason, in \paperii\ we investigated stationary states, the Thomas--Fermi approximation and the Bogoliubov excitation spectrum of our model of coupled atoms and molecules, and in fact found that even when
$a_s =890 a_0$, the corrections were quite modest, though quite perceptible. The Bragg scattering experiment is essentially a measurement of the excitation spectrum, and the frequency changes it presents are a measure of the deviation from the spectrum expected of the corresponding hard sphere model.  These corrections are in fact quite modest; only about 10\% of the actual Bogoliubov quasiparticle frequencies.

\subsection{Relating Theory to Experiment}
The experiment set out to test the limits of conventional theory, and convincingly achieved that aim. However the procedure used was not ideal for comparison with our detailed model.  The most challenging problem is the absence of any measurements of the initial state of the condensate immediately before applying the Bragg pulse.  The issue is further complicated by the procedure used to enhance the density of the condensate, before ramping the scattering length to the value used for the Bragg pulse.  The result is an initial state for the Bragg scattering which, not being a stationary state, cannot be definitively determined.  In order to compare our computations with experiment we have relied on the time and space averaged density measurements implicit in their presentation of the frequency shifts expected  from the Huang--Yang theory.  We have converted these to the appropriate values of the time-averaged density-weighted density, and using these we achieve our results, which are in very good quantitative agreement with the measured results.  

We would consider it of importance in any future experiments to present 
\begin{enumerate}
\item \emph{Either }: Measurements of the initial state;
\item \emph{Or }: A precise quantitative description of the procedure used to create each initial state from the initial condensate, which can be reliably modelled as a stationary Bose--Einstein condensate. 
\end{enumerate}
The presentation of results as time averaged quantities should be avoided; these create very significant computational difficulties.

\subsection{Further Opportunities}
The methods we have developed can clearly be applied to other problems in which the flexible adjustment of the scattering length afforded by Feshbach resonances has been exploited, for example the Bose-Nova problem, and the  related problem of bright solitons.  It is also conceivable that the methodology could be extended to study the physics of Efimov states in the presence of a Bose--Einstein condensate.


\section*{Acknowledgments}
The research in this paper was supported by 
the New Zealand Foundation for Research, Science and
Technology under Contract No. NERF-UOOX0703, ``Quantum
Technologies'' and Marsden Contract No. UOO509.

\appendix

\section{Simulation parameters}
\label{sec:parameters}
The values of different parameters used in the simulations are listed in this section.
The duration of the Bragg pulse is inferred from the data for the widths of the Bragg pulse (Fig. 3(b) in \cite{papp2008}), where the contribution from the pulse duration will be inversely proportional to the pulse length as $\Delta\omega = 0.36/t_{\text{pulse}}$. The duration for the different scattering lengths are listed in Tab.\,\ref{tab:App}, where we also list the values of the three-body loss parameter $\gamma$. The loss parameter has been determined by using the corresponding values of the three-body recombination rate $K_3$ given by Braaten \emph{et al.} \cite{braaten2007}.
\begin{table}[hbt]
\caption{\label{tab:App}Pulse lengths and three-body loss parameters for different scattering lengths. The data for the width are taken from Fig. 3 in \cite{papp2008}, and values of the loss parameter are calculated using values of the three-body recombination rate in \cite{braaten2007}.}
\begin{center}\small
\begin{tabular}{c|c|c|r}\hline
Scattering length&Width from duration &Pulse length &Three-body loss \\
 $a_s$ [$a_0$]&$\Delta f$ [kHz]&$t$ [ms]&$\gamma$ (cm$^{-3}$/s)  \\
\hline
150&0.8&0.45&$5\times10^{-29}$\\
300&1.5&0.24&$1.5\times10^{-27}$\\
500&2.5&0.14&$5\times10^{-27}$\\
585&3.0&0.12&$1.5\times10^{-26}$\\
700&3.4&0.11&$2\times10^{-26}$\\
800&3.7&0.10&$2.5\times10^{-26}$\\
890&4.0&0.09&$3\times10^{-26}$\\
\hline
\end{tabular}
\end{center}
\end{table}

\section{Projectors and momentum space truncation}

\subsection{Momentum space truncation}
\label{sec:AppTruncation}
To include all the physics that we are interested in, the momentum space needs to include at least the first order Bragg momentum, at $|\boldsymbol{k}| = |\boldsymbol{q}|$. Assuming the Bragg pulse is only applied in the $x$-direction, we can write the optical potential as
\begin{equation}
V_{\text{opt}} = V_{0}\cos{(Qx- \omega t)} = \frac{V_0}{2}(e^{i(Qx - \omega t)} +e^{-i(Qx - \omega t)}) \mbox{,}
\end{equation}
where $Q = |\boldsymbol{q}|$.

To fulfill this condition as well as the condition that the number of grid points $N_{text{grid}} = 2^{n}$ for some positive integer $n$, the number of grid points in the $x$-direction is chosen to be $N_x=2048$. The $y$- and $z$-directional grids are chosen to have $N_y = N_z = 64$, since these directions are of less importance, making the total number of grid points $2\times2048\times64\times64$. This would not only make the simulations very computationally heavy, but also, since in the c-field formalism we will have on average half a quasiparticle of noise per mode in the initial state, we would get many more noise particles than condensate particles. According to the validity condition for the c-field methods \cite{norrie2006}, this would make our simulations invalid.

To get around this problem but still include all the momentum space of relevance, we neglect the parts of momentum space where the population will be insignificant, and include only those modes that are initially populated or where we can expect to get significant population from scattering. 
Because the interest here is Bragg scattering with a Bragg pulse applied in the positive $x$-direction, we divide momentum space into bands in this direction, each centered around $nQ$ for some $n$, where $Q$ is the momentum of the pulse. We can  thus write the wavefunctions $\psi$ and $\phi$ as
\begin{align}
\psi\ofx &= \sum_n \psi_n\ofx e^{inQx} ,\\
\phi\ofx &= \sum_n \phi_n\ofx e^{inQx} ,
\end{align}
where $\psi_n$ and $\phi_n$ are the Fourier transforms of the momentum space wavefunction in the band centred around $nQ$.

The projectors $\mathcal{P}_a$ and $\mathcal{P}_m$ are given by
\begin{eqnarray}
\mathcal{P}_a &=& \Theta \left(\left(\frac{k_{x}^{2}}{k_{x,\text{cut}}^{2}} + \frac{k_y^2}{k_{y,\text{cut}}^2} + \frac{k_z^2}{k_{z,\text{cut}}^2} \right)- 1\right)  \\
\mathcal{P}_m &=& \Theta \left(\frac{1}{4}\left(\frac{k_{x}}{k_{x,\text{cut}}^{2}} + \frac{k_y^2}{k_{y,\text{cut}}^2} + \frac{k_z^2}{k_{z,\text{cut}}^2} \right)- 1\right)
\end{eqnarray}
with the $x$-directional cutoff now given by
\begin{equation}
k_{x,\text{cut}} = \frac{\Delta k}{4},
\end{equation}
where $\Delta k$ is the width of each momentum space band. The projectors are the same for all the wavefunctions $\psi_n$ and $\phi_n$; for each band the projectors are ellipsoids centered around the midpoint at $y=z=0$ and $x=nQ$. 

\subsubsection{Four significant bands}
\begin{figure}[t]
\begin{center}
\begin{psfrags}%
\psfrag{s01}[t][t]{\color[rgb]{0,0,0}\setlength{\tabcolsep}{0pt}\begin{tabular}{c}$k_x$ $[x_0^{-1}]$\end{tabular}}%
\psfrag{s02}[b][b]{\color[rgb]{0,0,0}\setlength{\tabcolsep}{0pt}\begin{tabular}{c}$k_y$ $[x_0^{-1}]$\end{tabular}}%
\psfrag{s03}[b][b]{\color[rgb]{0,0,0}\setlength{\tabcolsep}{0pt}\begin{tabular}{c}Atom projector\end{tabular}}%
\psfrag{s05}[t][t]{\color[rgb]{0,0,0}\setlength{\tabcolsep}{0pt}\begin{tabular}{c}$k_x$ $[x_0^{-1}]$\end{tabular}}%
\psfrag{s06}[b][b]{\color[rgb]{0,0,0}\setlength{\tabcolsep}{0pt}\begin{tabular}{c}$k_y$ $[x_0^{-1}]$\end{tabular}}%
\psfrag{s07}[b][b]{\color[rgb]{0,0,0}\setlength{\tabcolsep}{0pt}\begin{tabular}{c}Molecule projector\end{tabular}}%
\psfrag{s13}[lt][lt]{\color[rgb]{0,0,0}\setlength{\tabcolsep}{0pt}\begin{tabular}{l}$k_{x,\text{cut}}$\end{tabular}}%
\psfrag{s14}[lt][lt]{\color[rgb]{0,0,0}\setlength{\tabcolsep}{0pt}\begin{tabular}{l}$k_{y,\text{cut}}$\end{tabular}}%
\psfrag{s15}[lt][lt]{\color[rgb]{0,0,0}\setlength{\tabcolsep}{0pt}\begin{tabular}{l}$\Delta k$\end{tabular}}%
\psfrag{x01}[t][t]{0}%
\psfrag{x02}[t][t]{0.1}%
\psfrag{x03}[t][t]{0.2}%
\psfrag{x04}[t][t]{0.3}%
\psfrag{x05}[t][t]{0.4}%
\psfrag{x06}[t][t]{0.5}%
\psfrag{x07}[t][t]{0.6}%
\psfrag{x08}[t][t]{0.7}%
\psfrag{x09}[t][t]{0.8}%
\psfrag{x10}[t][t]{0.9}%
\psfrag{x11}[t][t]{1}%
\psfrag{x12}[t][t]{-10}%
\psfrag{x13}[t][t]{-5}%
\psfrag{x14}[t][t]{0}%
\psfrag{x15}[t][t]{5}%
\psfrag{x16}[t][t]{10}%
\psfrag{x17}[t][t]{15}%
\psfrag{x18}[t][t]{20}%
\psfrag{x19}[t][t]{-10}%
\psfrag{x20}[t][t]{-5}%
\psfrag{x21}[t][t]{0}%
\psfrag{x22}[t][t]{5}%
\psfrag{x23}[t][t]{10}%
\psfrag{x24}[t][t]{15}%
\psfrag{x25}[t][t]{20}%
\psfrag{v01}[r][r]{0}%
\psfrag{v02}[r][r]{0.1}%
\psfrag{v03}[r][r]{0.2}%
\psfrag{v04}[r][r]{0.3}%
\psfrag{v05}[r][r]{0.4}%
\psfrag{v06}[r][r]{0.5}%
\psfrag{v07}[r][r]{0.6}%
\psfrag{v08}[r][r]{0.7}%
\psfrag{v09}[r][r]{0.8}%
\psfrag{v10}[r][r]{0.9}%
\psfrag{v11}[r][r]{1}%
\psfrag{v12}[r][r]{-10}%
\psfrag{v13}[r][r]{-5}%
\psfrag{v14}[r][r]{0}%
\psfrag{v15}[r][r]{5}%
\psfrag{v16}[r][r]{10}%
\psfrag{v17}[r][r]{-10}%
\psfrag{v18}[r][r]{-5}%
\psfrag{v19}[r][r]{0}%
\psfrag{v20}[r][r]{5}%
\psfrag{v21}[r][r]{10}%
\resizebox{12cm}{!}{\includegraphics{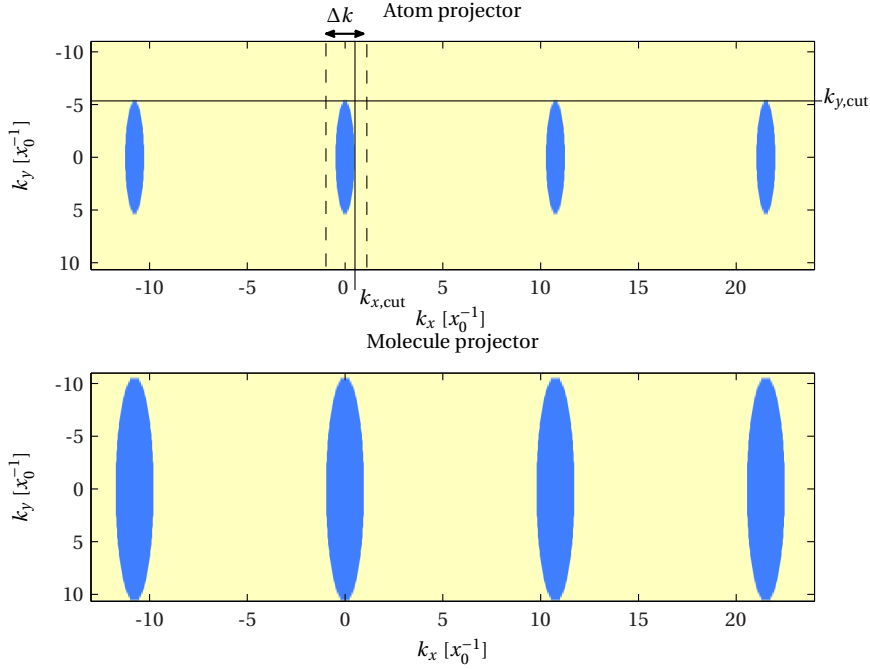}}%
\end{psfrags}%
\caption{Atom field projector (top panel) and molecule field projector (bottom panel) in the $xy$-plane for the case of four bands in momentum space being significantly populated. The width of each band is $\Delta k$. The blue areas indicate the regions of momentum space that the wavefunctions are projected into, determined by the parameters $k_{x,\text{cut}}$ and $k_{y,\text{cut}}$. The parameter $x_0$ is the length scale associated with the $x$-axis of the trap, given by $x_0 = \sqrt{\hbar/2m\omega_x}\approx6.665\times10^{-7}$ m. Here $\Delta k = 1.3\times10^{-6}$ m, $k_{x,\text{cut}}= 3.2\times10^{-7}$ m, and $k_{y,\text{cut}}= 3.5\times10^{-6}$ m. }
\label{fig:Projector}
\end{center}
\end{figure}

We find that only the four momentum bands
corresponding  to the orders  $n = -1,0,1,2$ will be significant populated during our simulations. We then have 
\begin{align}
\psi\ofx &= \psi_{-1}\ofx e^{-iQx} + \psi_0\ofx + \psi_1\ofx e^{iQx} + \psi_2\ofx e^{2iQx}  \\
\phi\ofx &=  \phi_{-1}\ofx e^{-iQx} + \phi_0\ofx + \phi_1\ofx e^{iQx} + \phi_2\ofx e^{2iQx} .
\end{align}
Fig.\,\ref{fig:Projector}, shows the projector for the full wavefunctions for this case in the $xy$-plane, where we have also indicated the width of each band $\Delta k$ and the momentum space cutoffs $k_{x,\text{cut}}$ and $k_{y,\text{cut}}$.

This gives us the following expression for the squared norm of $\psi$
\begin{eqnarray}
|\psi\ofx|^2 &=& A_{-3}\ofx e^{-iQx} +A_{-2}\ofx e^{-iQx} +A_{-1}\ofx e^{-iQx} + \nonumber \\
&& + A_0\ofx + A_1\ofx e^{iQx} + A_2\ofx e^{2iQx}
\end{eqnarray}
where 
\begin{eqnarray}
A_0 &=&  |\psi_{-1}|^2 + |\psi_{0}|^2 + |\psi_{1}|^2  +  |\psi_{2}|^2     \\
A_1 &=&  \psi^*_{-1}\psi_0 + \psi^*_{0}\psi_1  + \psi^*_{1}\psi_2 =A^*_{-1}    \\
A_2 &=&  \psi^*_{-1}\psi_1 + \psi^*_{0}\psi_2 = A^*_{-2} \\
A_3 &=&  \psi^*_{-1}\psi_2 = A^*_{-3} 
\end{eqnarray}
Similarly for the squared norm of $\phi$ we have
\begin{eqnarray}
|\phi\ofx|^2 &=& B_{-3}\ofx e^{-iQx} +B_{-2}\ofx e^{-iQx} +B_{-1}\ofx e^{-iQx} + \nonumber \\
&& + B_0\ofx + B_1\ofx e^{iQx} + B_2\ofx e^{2iQx}
\end{eqnarray}
where 
\begin{eqnarray}
B_0 &=&  |\phi_{-1}|^2 + |\phi_{0}|^2 + |\phi_{1}|^2  +  |\phi_{2}|^2     \\
B_1 &=&  \phi^*_{-1}\phi_0 + \phi^*_{0}\phi_1  + \phi^*_{1}\phi_2 =B^*_{-1}    \\
B_2 &=&  \phi^*_{-1}\phi_1 +  \psi^*_{0}\phi_2 = B^*_{-2} \\
B_3 &=&  \phi^*_{-1}\phi_2 = B^*_{-3} 
\end{eqnarray}

The density squared now becomes
\begin{eqnarray}
\left(|\psi\ofx|^2 +  |\phi\ofx|^2\right)^2 &=& C_{-3}\ofx e^{-3iQx} +C_{-2}\ofx e^{-2iQx} +C_{-1}\ofx e^{-iQx} + \nonumber \\
&& + C_0\ofx + C_1\ofx e^{iQx} + C_2\ofx e^{2iQx}+ C_3\ofx e^{3iQx} 
\end{eqnarray}
where 
\begin{eqnarray}
C_0 &=&  (A_{0}+2B_0)^2 + 2|A_{1}+2B_{1}|^2 + 2|A_{2}+2B_{2}|^2  + 2|A_{3}+2B_{3}|^2    \\
C_1 &=&  2(A_{0}+2B_0)(A_{1}+2B_{1}) + 2(A_{-1}+2B_{-1}) (A_{2}+2B_{2}) \nonumber\\
&& + 2(A_{-2}+2B_{-2})(A_{3}+2B_{3})  =C^*_{-1}    \\
C_2 &=& (A_{1}+2B_{1})^2 + 2(A_{0}+2B_0)(A_{2}+2B_{2}) + 2(A_{-1}+2B_{-1})(A_{3}+2B_{3})   = C^*_{-2} \\
C_3 &=& 2(A_{0}+2B_0)(A_{3}+2B_{3}) + 2(A_{1}+2B_{1}) (A_{2}+2B_{2})    = C^*_{-3} 
\end{eqnarray}
and terms with $n<-3$ or $n>3$ have been neglected. 

These expressions, together with the expressions for $\psi$ and $\phi$ are substituted into the equation of motion for the atom wave function (\ref{eq:dPsi_dt}). Dropping all terms with a factor of $e^{inQx}$ with $n\neq-1,0,1,2$, and collecting terms corresponding to the same band together, we get
\begin{eqnarray}
\rmi\hbar\frac{\partial\psi_{-1}}{\partial t} &=&-\frac{\hbar^2\tilde\nabla_{-1}^{2}}{2m}{\psi_{-1}} + \mathcal{P}_a\left\{V_{a}\psi_{-1} +\frac{V_0}{2}\psi_{0} e^{i\omega t} \right. \nonumber \\
&&+ U_{aa}\left(A_0\psi_{-1}  + A_{-1}\psi_{0} + A_{-2}\psi_{1}+ A_{-3}\psi_{2}  \right) +g\left(\psi_{0}^*\phi_{-1} +\psi_1^*\phi_0  + \psi_{2}^*\phi_{1}  \right)    \nonumber\\
&& \left.  - i\gamma\left(C_0\psi_{-1}  + C_{-1}\psi_{0} + C_{-2}\psi_{1} + C_{-3}\psi_{2} \right) \right\} \\
\rmi\hbar\frac{\partial\psi_0}{\partial t} &=& -\frac{\hbar^2\tilde\nabla^{2}_0}{2m}{\psi_0} + \mathcal{P}_a\left\{V_{a}\psi_0 +\frac{V_0}{2}\left(\psi_{-1} e^{-i\omega t} + \psi_1 e^{i\omega t} \right) \right. \nonumber \\
&& + U_{aa}\left(A_{1}\psi_{-1} + A_0\psi_{0} + A_{-1}\psi_{1} + A_{-2}\psi_{2} \right) +g\left(\psi_{-1}^*\phi_{-1} +\psi_0^*\phi_0 + \psi_1^* \phi_1   + \psi_2^* \phi_2  \right)\nonumber\\
&& \left.   - i\gamma\left(C_{1}\psi_{-1}  + C_0\psi_{0} +  C_{-1}\psi_{1} +C_{-2}\psi_{2}\right) \right\} \\
\rmi\hbar\frac{\partial\psi_1}{\partial t} &=&-\frac{\hbar^2\tilde\nabla_1^{2}}{2m}{\psi_1} + \mathcal{P}_a\left\{V_{a}\psi_1 +\frac{V_0}{2}\left(\psi_{0} e^{-i\omega t} + \psi_2 e^{i\omega t} \right)\right. \nonumber \\
&&+ U_{aa}\left(A_{2}\psi_{-1}+ A_{1}\psi_{0} + A_0\psi_{1}  + A_{-1}\psi_{2} \right) \nonumber\\
&&  \left. +g\left(\psi_{1}^*\phi_{2} +\psi_0^*\phi_1 + \psi_{-1}^* \phi_0  \right) - i\gamma\left(C_0\psi_{1}  + C_{1}\psi_{0} + C_{2}\psi_{-1} \right) \right\} \\
\rmi\hbar\frac{\partial\psi_2}{\partial t} &=& -\frac{\hbar^2\tilde\nabla_2^{2}}{2m}{\psi_2} + \mathcal{P}_a\left\{V_{a}\psi_2 +\frac{V_0}{2}\psi_{1} e^{-i\omega t} + U_{aa}\left(A_3\psi_{-1} + A_{2}\psi_{0} + A_{1}\psi_{1} + A_{0}\psi_{2} \right) \right. \nonumber\\
&& \left. +g\left(\psi_{-1}^*\phi_{1} +\psi_0^*\phi_2 \right) - i\gamma\left(C_3\psi_{-1}  + C_{2}\psi_{0} + C_{1}\psi_{1} + C_{0}\psi_{2} \right) \right\}, 
\end{eqnarray}
where 
\begin{align}
\tilde\nabla_n^2 = \nabla^2 + i2n\frac{\partial}{\partial x} - n^2Q^2 .
\end{align}

Similarly, the equations of motion for the bands of the molecule wavefunction (\ref{eq:dPhi_dt}), become
\begin{eqnarray}
\rmi\hbar\frac{\partial\phi_{-1}}{\partial t} &=& -\frac{\hbar^2\tilde\nabla_{-1}^{2}}{4m}{\phi_{-1}} + \mathcal{P}_m\left\{V_{m}\phi_{-1} +V_0\phi_{0} e^{i\omega t} + {g}\psi_{-1}\psi_{0}  \right\} \\
\rmi\hbar\frac{\partial\phi_{0}}{\partial t} &=& -\frac{\hbar^2\tilde\nabla_0^{2}}{4m}{\phi_{0}} + \mathcal{P}_m\left\{V_{m}\phi_{0} +{V_0}\left(\phi_{-1} e^{-i\omega t} + \phi_1 e^{i\omega t} \right) \right.\nonumber\\
&&+ \left.\frac{g}{2}\left(2\psi_{-1}\psi_{1} + \psi_0^2 \right)\right\} \\
\rmi\hbar\frac{\partial\phi_{1}}{\partial t} &=& -\frac{\hbar^2\tilde\nabla_1^{2}}{4m}{\phi_{1}} + \mathcal{P}_m\left\{V_{m}\phi_{1} +{V_0}\left(\phi_{0}  e^{-i\omega t} + \phi_2  e^{i\omega t} \right) \right.\nonumber\\
&&\left.+ g\left(\psi_{-1}\psi_{2} + \psi_0\psi_1 \right)\right\} \\
\rmi\hbar\frac{\partial\phi_{2}}{\partial t} &=& -\frac{\hbar^2\tilde\nabla_2^{2}}{4m}{\phi_{2}} + \mathcal{P}_m\left\{V_{m}\phi_{2} +{V_0}\phi_{1} e^{-i\omega t}  + \frac{g}{2}\left(2\psi_0\psi_2+ \psi_1^2 \right)\right\}. 
\end{eqnarray}

\subsection{Renormalization parameter}
\label{sec:AppRenormalization}
In order to calculate the renormalization constant, we need to evaluate the integral
\begin{equation}
\mathcal{I}=\int_V{\frac{d\bfk}{k^2}} ,
\end{equation}
where the relationship between the renormalization factor, $\Lambda$, and the integral is $\mathcal{I} = 4\pi\Lambda$.

In the simplest case, the momentum space cutoff is the same in all directions, and the volume of the populated low energy subspace is spherical. In this case we can use spherical coordinates, to get
\begin{equation}
\mathcal{I} = \int^{k_{R,\text{cut}}}_0 \int^{2\pi}_0 \int^\pi_0 \frac{\rho^2 d\theta d\varphi d\rho}{\rho^2} = \int^{k_{R,\text{cut}}}_0{4\pi}d\rho =4\pi {k_{R,\text{cut}}},
\end{equation}
where $k_{R,\text{cut}}$ is the value of the cutoff, so that we get the simple relationship between the renormalization factor and the cutoff $\Lambda = {k_{R,\text{cut}}}$.

\subsubsection{Anisotropic cutoff}
In the case of an anisotropic momentum space cutoff, calculating the renormalization factor becomes slightly more complicated.
In our simulations we have, as is described in section \ref{sec:AppTruncation}, a momentum space that is divided into four bands. We therefore have to calculate
\begin{equation}
\mathcal{I} = \sum_n \mathcal{I}_n = \sum_n\int_{V_n}{\frac{d\bfk}{k^2}}
\end{equation}
where $V_n$ is the volume of the low energy subspace in band $n$. 

Each band in the truncation has an ellipsoidal projector, symmetric in the $yz$-plane, with maximum value $k_{y,\text{cut}} = k_{z,\text{cut}} \equiv k_{yz,\text{cut}}$. In the $x-$direction, each band is centered around $k_x = nQ$ and they all have the same width $\Delta k$. 

Because of the cylindrical symmetry of the volume we can simply the problem by changing to polar coordinates to get
\begin{eqnarray}
\mathcal{I}_n &=& \int^{nQ+\Delta k/2}_{nQ-\Delta k/2}\int^{\rho_\text{max}(\zeta)}_0 \frac{2\pi \rho}{\rho^2+\zeta^2}d\rho d\zeta \\
&=& 2\pi \int^{nQ+\Delta k/2}_{0}\left(\ln\left(\rho^2_{max}(\zeta)+\zeta^2\right) - \ln\left(\zeta^2\right)\right)d\zeta 
\end{eqnarray}
where $\rho_\text{max}$ is given by 
\begin{equation}
\rho^2_\text{max}(\zeta) = k_{yz,\text{cut}}^2\left(1-\frac{\zeta^2}{(nQ+\Delta k/2)^2}\right).
\end{equation}
We then get
\begin{equation}
\mathcal{I}_n = 2\pi \int^{\Delta k/2}_{0}\left(\ln\left(\zeta^2\left(1-\frac{k_{yz,\text{cut}}^2}{(\Delta k/2)^2}\right)+2nQ\zeta+(nQ)^2+k_{yz,\text{cut}}^2\right)-2\ln\left(\zeta+nQ\right)\right)d\zeta .
\end{equation}

Since the volume $V_n$ is an ellipsoid and not a sphere, we have that $ k_{yz,\text{cut}}^2\neq (\Delta k/2)^2$, and since $\Delta k<2Q$, i.e. the bands are not overlapping, this integral has the solution\begin{eqnarray}
\mathcal{I}_n &=& 2\pi \sqrt{\frac{2nQ}{1/\tau({k_{R,\text{cut}}})-1}-2{k^2_{R,\text{cut}}}} \nonumber \\
&& \times \ln\left(\frac{\frac{\Delta k}{2}\left(2nQ-2{k_{R,\text{cut}}}\sqrt{1-\tau({k_{R,\text{cut}}})-\tau(nQ)} \right)+4\left(1-\tau({k_{R,\text{cut}}})\right)\left((nQ)^2-{k^2_{R,\text{cut}}}\right)}{\frac{\Delta k}{2}\left(2nQ+2{k_{R,\text{cut}}}\sqrt{1-\tau({k_{R,\text{cut}}})-\tau(nQ)} \right)+4\left(1-\tau({k_{R,\text{cut}}})\right)\left((nQ)^2-{k^2_{R,\text{cut}}}\right)} \right) \nonumber \\
&& + \frac{4\pi nQ}{1 - \tau({k_{R,\text{cut}}})}\ln\left(\frac{nQ+\Delta k/2}{\sqrt{(nQ)^2+{k^2_{R,\text{cut}}}}} \right) - 2\pi nQ\ln\left(nQ\left(nQ+\Delta k/2\right)\right)   ,
\end{eqnarray}
where $\tau(x) = x^2/(\Delta k/2)^2$.

\section{Density-weighted density}
\label{sec:AppDensity}
We wish to calculate the density-weighted density $\overline{n(t)}$ for our coupled atom and molecule system in the Wigner formalism used in this paper.

\subsection{Wigner ordering}
For an operator $\hat{a}$, we know the symmetrically ordered average,
\begin{eqnarray}
\left\{\hat{N}^2\right\}_{\text{sym}} &\equiv& \left\{\hat{a}^2\hat{a}^{\dagger2}\right\}_{\text{sym}} \nonumber \\
&=&\frac{1}{6}\left\{\hat{a}^2\hat{a}^{\dagger2}+\hat{a}^{\dagger}\hat{a}\hat{a}^{\dagger}\hat{a}+\hat{a}\hat{a}^{\dagger2}\hat{a}+\hat{a}\hat{a}^{\dagger}\hat{a}\hat{a}^{\dagger}+\hat{a}^{\dagger2}\hat{a}^2\right\}.
\end{eqnarray}
Assuming that the commutator is
\begin{equation}
\left[\hat{a},\hat{a}^\dagger\right] = \Delta,
\end{equation}
and that 
\begin{equation}
N = \hat{a}^\dagger\hat{a},
\end{equation}
we find 
\begin{equation}
\left\{\hat{N}^2\right\}_{\text{sym}} = N^2+2\Delta N +\frac{\Delta^2}{2}.
\end{equation}
Since we also have
\begin{equation}
\left\{\hat{N}\right\}_{\text{sym}} = N +\frac{\Delta}{2},
\end{equation}
we get
\begin{equation}
N^2 = \left\{\hat{N}^2\right\}_{\text{sym}} - 2\Delta\left\{\hat{N}\right\}_{\text{sym}} + \frac{\Delta^2}{2}.
\end{equation}
We therefore get the averages
\begin{eqnarray}
\overline{N} &=& \left\langle \left\{\hat{N}\right\}_{\text{sym}} - \frac{\Delta}{2} \right\rangle, \\
\overline{N^2} &=& \left\langle \left\{\hat{N}^2\right\}_{\text{sym}} - 2\Delta\left\{\hat{N}\right\}_{\text{sym}} + \frac{\Delta^2}{2} \right\rangle.
\end{eqnarray}

\subsection{Atom-molecule density-weighted density}
We now consider the case of an atom operator $\hat\psi(\x,t)$ and a molecule operator $\hat\phi(\x,t)$, with commutators
\begin{eqnarray}
\left[\hat{\psi}(\x,t),\hat{\psi}^\dagger(\x,t)\right] &=& \Delta_a, \\
\left[\hat{\phi}(\x,t),\hat{\phi}^\dagger(\x,t)\right] &=& \Delta_m,
\end{eqnarray}
The average total atom number for this system is given by
\begin{equation}
\overline{N(t)}  = \left\langle \int{d\x\left[\hat{\psi}^\dagger(\x,t)\hat{\psi}(\x,t) + 2\hat{\phi}^\dagger(\x,t)\hat{\phi}(\x,t)\right]} \right\rangle,
\end{equation}
where, as usual, we count a molecule as two atoms.
Using the commutation relations this can be expressed as
\begin{equation}
\overline{N(t)} =  \left\langle \int{d\x\left(\left\{\hat{n}_a(\x,t)\right\}_{\text{sym}} + 2\left\{\hat{n}_m(\x,t)\right\}_{\text{sym}} - \frac{\Delta_a}{2} - \Delta_m\right) } \right\rangle,
\end{equation}
where $\left\{\hat{n}_a(\x,t)\right\}_{\text{sym}}$ and $\left\{\hat{n}_m(\x,t)\right\}_{\text{sym}}$ are the symmetrically ordered averages
\begin{eqnarray}
\left\{\hat{n}_a(\x,t)\right\}_{\text{sym}} &\equiv& \left\{\hat\psi^\dagger(\x,t)\hat\psi(\x,t)\right\}_{\text{sym}}, \\
\left\{\hat{n}_m(\x,t)\right\}_{\text{sym}} &\equiv& \left\{\hat\phi^\dagger(\x,t)\hat\phi(\x,t)\right\}_{\text{sym}}. 
\end{eqnarray}

The density-weighted density for the system is given by
\begin{equation}
\overline{n(t)}=\frac{1}{\overline{N(t)}}\left\langle\int{d\x\left(\hat{\psi}^\dagger(\x,t)\hat{\psi}(\x,t) + 2\hat{\phi}^\dagger(\x,t)\hat{\phi}(\x,t)\right)^2}\right\rangle. 
\end{equation}
Using the same approach as in the previous section, we can express this as
\begin{eqnarray}
\overline{n(t)}&=& \frac{1}{\overline{N(t)}} \left\langle \int d\x\left(\overline{n^2_a(\x,t)} + 4\overline{n^2_m(\x,t)} - 2\Delta_m\left\{\hat{n}_a(\x,t)\right\}_{\text{sym}} -  2\Delta_a\left\{\hat{n}_m(\x,t)\right\}_{\text{sym}} \right. \right. \nonumber \\
&& \left.\left. + 4 \left\{\hat{n}_a(\x,t)\right\}_{\text{sym}} \left\{\hat{n}_m(\x,t)\right\}_{\text{sym}}  +\Delta_a\Delta_m \right) \right\rangle,
\end{eqnarray}
where $\overline{n^2_a(\x,t)}$ and $\overline{n^2_m(\x,t)}$ are given by
\begin{eqnarray}
\overline{n^2_a(\x,t)} &=& \left\{\hat{n}^2_a(\x,t)\right\}_{\text{sym}} -2\Delta_a\left\{\hat{n}_a(\x,t)\right\}_{\text{sym}}+\frac{\Delta_a^2}{2}, \\
\overline{n^2_m(\x,t)} &=& \left\{\hat{n}^2_m(\x,t)\right\}_{\text{sym}} -2\Delta_m\left\{\hat{n}_m(\x,t)\right\}_{\text{sym}}+\frac{\Delta_m^2}{2}, 
\end{eqnarray}
where 
\begin{eqnarray}
\left\{\hat{n}^2_a(\x,t)\right\}_{\text{sym}} &\equiv& \left\{\hat\psi^{\dagger2}(\x,t)\hat\psi^2(\x,t)\right\}_{\text{sym}}, \\
\left\{\hat{n}^2_m(\x,t)\right\}_{\text{sym}} &\equiv& \left\{\hat\phi^{\dagger2}(\x,t)\hat\phi^2(\x,t)\right\}_{\text{sym}}. 
\end{eqnarray}

\subsection{Check with initial state}
The initial state corresponds to the two states $\hat\psi\ofx$ and $\hat\phi\ofx$, given by
\begin{eqnarray}
\hat\psi\ofx &=& \psi_0\ofx + \frac{\hat{r}\ofx}{\sqrt{2}}, \\
\hat\phi\ofx &=& \phi_0\ofx + \frac{\hat{s}\ofx}{\sqrt{2}},
\end{eqnarray}
where
\begin{eqnarray}
\left\langle|\hat{r}\ofx|^2\right\rangle &=& \Delta_a, \\
\left\langle|\hat{r}\ofx|^4\right\rangle&=& 2\Delta_a^2, \\
\left\langle|\hat{s}\ofx|^2\right\rangle &=& \Delta_m, \\
\left\langle|\hat{s}\ofx|^2\right\rangle &=& 2\Delta_m^2, 
\end{eqnarray}
Then we have
\begin{eqnarray}
\left\langle \left\{\hat{n}^2_a\ofx\right\}_{\text{sym}} \right\rangle &=& \left\langle \left| \psi_0\ofx +\frac{\hat{r}\ofx}{\sqrt{2}}\right|^4\right\rangle \nonumber \\
&=& \left|\psi_0\ofx\right|^4 + 2\Delta_a\left|\psi_0\ofx\right|^2 +\frac{\Delta_a^2}{2}, \\
\left\langle \left\{\hat{n}_a\ofx\right\}_{\text{sym}} \right\rangle &=& \left|\psi_0\ofx\right|^2 +\frac{\Delta_a}{2}, \\
\left\langle \left\{\hat{n}^2_m\ofx\right\}_{\text{sym}} \right\rangle &=& \left\langle \left| \phi_0\ofx +\frac{\hat{s}\ofx}{\sqrt{2}}\right|^4\right\rangle \nonumber \\
&=& \left|\phi_0\ofx\right|^4 + 2\Delta_m\left|\phi_0\ofx\right|^2 +\frac{\Delta_m^2}{2}, \\
\left\langle \left\{\hat{n}_m\ofx\right\}_{\text{sym}} \right\rangle &=& \left|\phi_0\ofx\right|^2 +\frac{\Delta_m}{2}, \\
\left\langle \left\{\hat{n}_a\ofx\right\}_{\text{sym}} \left\{\hat{n}_m\ofx\right\}_{\text{sym}} \right\rangle &=& \left\langle \left\{\hat{n}_a\ofx\right\}_{\text{sym}} \right\rangle\left\langle \left\{\hat{n}_m\ofx\right\}_{\text{sym}} \right\rangle \nonumber \\
&=& \left|\psi_0\ofx\right|^2\left|\phi_0\ofx\right|^2 + \frac{\Delta_m\left|\psi_0\ofx\right|^2}{2} \nonumber \\
&& +\frac{\Delta_a\left|\phi_0\ofx\right|^2}{2} +\frac{\Delta_a\Delta_m}{4}.
\end{eqnarray}
And therefore
\begin{equation}
\overline{n} =  \frac{\int{d\x[|\psi_0\ofx|^2+2|\phi_0\ofx|^2]^2} }{\int{d\x[|\psi_0\ofx|^2+2|\phi_0\ofx|^2]}},
\end{equation}
as expected.

\bibliographystyle{unsrt}
\bibliography{bibuppsats}
\end{document}